\documentclass[%
reprint,
superscriptaddress,
amsmath,
amssymb,
aps,
floatfix,
]{revtex4-2}
\usepackage{xcolor}
\usepackage{graphicx}
\usepackage{dcolumn}
\usepackage{bm}
\usepackage{amssymb}
\usepackage{braket}
\usepackage[hidelinks]{hyperref}
\usepackage{subfigure}
\usepackage{outlines}
\usepackage{float}

\newcommand{\be}{\begin{equation} }
\newcommand{\p}{\partial}

\newcommand{\dtwomin}{$D^{2}_{min}$}
\newcommand{\xiol}{\xi/\mathcal{L}}
\newcommand{\ee}{\end{equation}}

\newcommand{\mlm}[1]{\textcolor{black}{#1}}
\newcommand{\ag}[1]{\textcolor{black}{#1}}

\begin{document}

\preprint{APS/123-QED}

\title{Yielding in dense active matter}

\author{Adil Ghaznavi}
\affiliation{Department of Physics and BioInspired Institute, Syracuse University, Syracuse, New York 13244, USA}

\author{Saverio Rossi}
\affiliation{Dipartimento di Fisica, Sapienza Università di Roma, Rome 00185 Italy}

\author{Francesco Zamponi}
\affiliation{Dipartimento di Fisica, Sapienza Università di Roma, Rome 00185 Italy}

\author{M. Lisa Manning}
\affiliation{Department of Physics and BioInspired Institute, Syracuse University, Syracuse, New York 13244, USA}
\email{mmanning@syr.edu}

\date{\today}

\begin{abstract}
High-density granular active matter is a useful model for dense animal collectives and could be useful for designing reconfigurable materials that can flow or solidify on command. Recent work has demonstrated key similarities and differences between the mechanical response of dense active matter and its sheared passive counterpart, yet a constitutive law that predicts precisely how dense active matter flows or fails remains elusive.  Here we study the yielding transition in dense active matter in the limit of slow driving and large persistence times, across a wide range of material preparations. Under shear, materials prepared to be very low energy or `ultrastable' are brittle, and well-described by elastoplastic constitutive laws.
We show that under random active forcing, however, ultrastable materials are always ductile. We develop a modified elastoplastic model that captures and explains these observations, where the key parameter is the correlation length of the input active driving field. We also observe large parameter regimes where the plastic flow is surprisingly well-predicted by the input active driving field and not highly dependent on the structural disorder, suggesting new strategies for control.
\end{abstract}

\maketitle


\section{\label{sec:level1}Introduction}
The framework of active matter, where energy is injected at the scale of individual agents, has been successfully used to describe collectives of animals and cells in nature. Examples include bacterial colonies~\cite{pierce_hydrodynamic_2018, dell2018growing_bacterial}, biological tissues~\cite{schotz_glassy_2013, bi_motility-driven_2016, yeomans2025active, kempf2019active}, starfish oocytes~\cite{foster2022active}, animal herds or flocks~\cite{cavagna_empirical_2010, cavagna2014bird, cavagna2015flocking, garcimartin2015flow}, and even human crowds~\cite{bottinelli2016emergent, gu2025emergence}. Such collectives exhibit interesting emergent behaviors, including phase transitions~\cite{tonertu1995XY} that cannot occur in non-active statistical mechanics descriptions.  

Active matter theories have also been useful from a materials design standpoint, where researchers have developed synthetic active systems~\cite{bililign2022motile, cohen2014emergent, cohen2014galvanotactic} to harness these emergent behaviors to drive new types of functionality. A particular subclass of active materials with non-reciprocal interactions~\cite{nonrecippopa2014, fruchart2020phase} can lead to macroscopic exotic ``odd elastic''~\cite{fruchart2023odd, ishimoto2023odd} or ``odd viscous'' behaviors, where shear forces can drive chiral rotations or vice versa. 

In order to predict how biological systems tune their function, and to engineer active materials with life-like adaptability, we need to understand the fundamental mechanisms that govern collective behavior in active matter at high densities, where agents are interacting with cages of their neighbors.  Emergent phenomena like flocking~\cite{tonertu1995XY, toner2018swarming, tonertu1998flocking, TONERturamaswamy2005170flocks, cavagna_empirical_2010, cavagna2014bird, cavagna2015flocking} and motility induced phase separation~\cite{wyart2014discontinuous} that occur at low-to-intermediate densities are relatively well understood, and can often be described by hydrodynamic theories~\cite{marchetti_hydrodynamics_2013}. However, dense active materials are dominated by steric interactions and therefore share many features of glass and jamming transitions~\cite{janssen2019active, henkes_active_2011}. This analogy has been the basis for ongoing work to understand how activity alters the glass transition~\cite{berthier2013non, berthier2017active, paul2023dynamical, nandi2018random} and short-time system dynamics, such as velocity autocorrelation functions~\cite{szamel2021long, henkes2020dense}. \mlm{In particular, very recent work has demonstrated a crossover from glass to jamming transitions as a function of the  persistence time of the active force~\cite{pareek2026}, and identified key features of the jammed phase in the infinitely persistent limit~\cite{gandikota2026jammed}.} 


An emerging related topic of interest is the rheology, plasticity, and yielding of dense active matter~\cite{berthier2025yielding}. This is a challenging topic because these systems inherit all of the difficulties of yielding in granular/glassy materials, such as plasticity, avalanches, intermittency, preparation and history dependence~\cite{mandal2020extreme, mandal2021rheology, morse_direct_2021}. Despite these difficulties, there have been useful theoretical and numerical advances, including a trap model that predicts particle displacements on both short and long timescales~\cite{woillez2020active}, methods for simulating dense active matter with very large persistence times~\cite{mandal2020multiple, mandal2021study}, a scaling analysis of the rheology near the jamming transition~\cite{liao2018criticality}, and a phase diagram of dynamical states in a sheared dense active system~\cite{mandal2021rheology}. 

Recent work builds on this to develop a continuum model~\cite{ghosh2025elastoplastic} that captures the steady-state behavior, including glassy relaxation and dynamical heterogeneties, in the presence of fluctuations.  However, a macroscopic theory (or constitutive law) that can predict the transition from an initial solid state to a flowing state in dense active matter -- i.e. yielding behavior, avalanche statistics, and flow patterns -- remains elusive. This restricts our ability to control and program behavior in dense active materials. 

Recent advances in understanding the yielding and flow of passive dense amorphous matter suggest it is a good time to revisit this question. A unifying framework for yielding has emerged, where the key parameter that tunes the yielding behavior and avalanche statistics is the disorder in the initial material preparation~\cite{berthier2025yielding}. Materials with high initial disorder or fluidity undergo ductile deformation, characterized by a shear stress that progressively increases with strain until reaching a steady-state plateau, and deformation occurring uniformly throughout the material volume. In contrast, materials with low initial disorder or fluidity exhibit brittle failure, with a large stress peak followed by a steep drop before reaching steady state, and deformation localized to distinct shear bands. This is termed the `ductile-to-brittle transition'.

There are several macroscopic constitutive theories that capture these features, including Elasto-Plastic Models (EPM)~\cite{nicolas_deformation_2018} and fluidity models~\cite{olmsted2008perspectives, fieldingcates2011nonlinear, catesfielding2006rheology, fielding2007complex}. Building on theories for Shear Transformation Zone (STZ)~\cite{falk1998dynamics} and Soft Glassy Rheology (SGR)~\cite{sollich2006soft}, both elasto-plastic and fluidity models assume that mesoscopic units of the material each possess a characteristic yield stress or resistance to flow, and that these regions are mechanically coupled, so that when a region flows it releases stress (via a Green's function) that affects other mesoscopic regions.  These models have been incredibly successful in describing, 
for example, the statistics of avalanches~\cite{coussot2002avalanche, baro2018universalavalanche, dahmen2011simpleavalanche, castellanos2018avalanche, karimi2017inertiaavalanche}. They also predict and explain precisely how the statistics of the mesoscopic yield stresses or fluidities control the `ductile-to-brittle' transition \cite{Ozawa2018,Popovi2018, pollard_yielding_2022,Rossi2022}. 

An unresolved question is whether this `ductile-to-brittle' transition is in the Random-Field Ising Model (RFIM) universality class, i.e., whether there is a true thermodynamic phase transition as a function of the disorder. EPMs exhibit many features of the AQS-driven RFIM transitions~\cite{Ozawa2018,Rossi2022}, although the long-range interactions via a non-positive definite Green's function so far obstructs an analytic proof, while fluidity models instead generate a crossover behavior with no true phase transition~\cite{Ozawa2018}. So far, it is hard to distinguish between the two scenarios in simulations~\cite{fieldingdivoux2024ductile, fieldingbarlow2020ductile,Rossi2022,Rossi2023}.

One of the reasons that it is difficult to distinguish between models for yielding is that the driving field is quite simple. Many simulations investigate athermal quasi-static shear (AQS) with Lees-Edwards boundary conditions that control the strain~\cite{maloney_amorphous_2006}; in linear response this is equivalent to a body force on each particle that points along the same axis with a linearly varying magnitude. Other simulations and experiments study yielding at finite strain rates~\cite{Salerno2012, nicolas2014, lagogianni2018}, though the applied field is still uniform shear.  Still others study the effect of an applied fixed force or stress, which are generally called creep experiments. These uniform strain driving fields suggest the mesoscopic regions all experience the same mean-field environment.

In contrast, self-propelled particle systems can be thought of as systems where there is an arbitrary field applied to each individual particle. In many simulations, this is modeled as an active self-propulsive force, which is a random version of a creep experiments~\cite{vicsek_novel_1995, henkes_active_2011}. An alternative approach is to study the effect of an applied active displacement  or strain, which is a random version of the strain-controlled system. Although in general active particles vary their direction of self-propulsion over a characteristic timescale $\tau$, it is useful to study the athermal, quasi-static limit where the direction of self-propulsion remains constant, \ag{i.e. the infinite persistence time limit,} termed Athermal Quasi-static Random Displacement (AQRD)~\cite{morse_direct_2021}, which is the random equivalent to the well-studied AQS. Previous work in the pre-yielding regime (the initial part of the stress-strain curve where the stress is still increasing with strain) demonstrate that there is a simple, direct relationship between the small avalanches that occur in sheared and active matter systems~\cite{morse_direct_2021}, predicted by infinite-dimensional mean field theory~\cite{agoritsas_mean-field_2021}. 

Additionally, mean field theory predicts that the dynamics for shear and random fields should remain the same up to a scaling factor throughout yielding~\cite{agoritsas_mean-field_2021}. Is this correct in two dimensions?

Here, we study the ductile-to-brittle yielding transition in dense active matter using both simulations and a modified elasto-plastic model. \ag{Recent work \cite{wiese2025avalanches} studies the nature of avalanches when a system is subjected to both shear \textit{and} activity. In our simulations we restrict ourselves to one method of driving at a time.} Our goals are two-fold; we wish to develop a predictive theory for the dynamics and emergent behavior of dense active matter subject to different initial conditions, which will be useful in understanding and controlling biological and engineered material systems. We also wish to use the active field as a new and different way to probe the physics of the yielding transition, allowing us to carefully test theories that make predictions for sheared {\it and} active matter, ultimately improving constitutive laws for such systems.

\section{Methods}

\subsection{Numerical Simulations of Dense Active Matter}
We adopt the breathing particle method~\cite{brito_theory_2018, kapteijns2019fast_breathing, ninarello2017models_breathing} to generate initial conditions for active matter simulations. Bidisperse packings are initialized and the radii are allowed to vary, which has been shown to generate highly stable packings~\cite{brito_theory_2018, wyartlernerdegiuli2015theory, Hagh_2022, fieldingdivoux2024ductile}. The ratio of the radii upon initialization is 1 to 1.4, to avoid crystallization~\cite{OHern2003}. The forces on the particles are given by a Hertzian interaction potential \cite{johnson1987contact}, modified with a spring potential term for the energy cost of radius variation,
\be
U=\sum_{ij}\Theta(\epsilon_{ij})\frac{\epsilon_{ij}^{5/2}}{5/2}+\sum_i \frac{k_\lambda}{2}\left(\frac{\lambda_i^0}{\lambda_i}\right)^2[(\lambda_i - \lambda_i^0)^2-\frac{1}{4}]\ ,
\ee
where $\Theta$ is the Heaviside function, $\epsilon_{ij}$ is the dimensionless particle overlap, $\lambda_i^0$ is the initial radius, $\lambda_i$ is the varying radius for particle $i$, and $k_\lambda$ is the effective spring constant describing resistance to changes in particle radii. The radial stiffness $k_\lambda$ controls the stability of the packing; small values of $k_\lambda$ generate more highly stable and ordered packings, as the system is given more freedom to find lower energy states.

We use athermal quasistatic random displacement (AQRD) \cite{morse_direct_2021} to generate the dynamics in our dense active matter system. We apply a displacement field $\Vec{c}$ to the system, normalized such that $|\Vec{c}|=1$, and move each particle a distance $u_i$ along the direction $c_i$. 

We also vary the correlation length of this input driving field. In standard self-propelled particle systems, the direction of self-propulsion is uncorrelated between each particle, with a correlation length equal to the particle size, $\xi =1$. 
However, in some flocking or nematic systems, the self-propulsion can become correlated over larger distances $\xi > 1$, though with periodic boundary conditions this lengthscale is restricted to $1/4$ of the box size. 
In AQS with Lees-Edwards boundary conditions, the lengthscale of the input driving field is $1/2$ the box size.  
We use Gaussian correlated random fields to implement $\Vec{c}$ with different correlation lengths as described previously~\cite{morse_direct_2021}, see Appendix~\ref{A:stress_def_aqrd} for more details.


After each small displacement, the system is allowed to find a new minimum subject to the constraint that it cannot move back along the direction $\Vec{c}$. This is implemented using a modified FIRE algorithm \cite{morse_direct_2021} that imposes a Lagrange multiplier preventing motion along $\Vec{c}$. We simulate each system until it has reached a strain that is equivalent to 20\% strain in AQS. This is sufficient to drive the system through the yielding transition, which typically occurs at about 5\% strain~\cite{richard_predicting_2020}. See Appendix~\ref{A:stress_def_aqrd} for details. 



\subsection{Analysis of numerical simulations}
\label{sec:analysis_simulation}
When dense active matter systems become unstable, particles rearrange. The disordered deformation field, which we will sometimes refer to as the `output' field, is quantified using a standard measure of the nonaffine deformation, \dtwomin~\cite{falk1998dynamics}.

We employ persistent homology~\cite{otter2017roadmap_persistenthomology}, a topological data analysis technique, to identify prominent clusters of particles with high values of plasticity (or any other metric), while ensuring that nearby, non-prominent clusters are merged. See Appendix~\ref{app:clusters} for details.

We are also interested in the capacity of the input field, $\Vec{c}$, to induce local strains. To quantify this, we study the best-fit affine deformation matrix, which is computed analogously to \dtwomin on a neighborhood of size $5 \bar{r}$ around each particle, where $\bar{r}$ is the mean particle radius.
Specifically, we choose $\Vec{X}_1$ to be the initial configuration of our solid, and $\Vec{X}_2 = \Vec{X}_1 + \Vec{c}$. We find $\epsilon$, the best fit affine transformation from $\Vec{X_1}$ to $\Vec{X_2}$. Then the total deviatoric strain in a neighborhood around each particle is given by $\tilde{\epsilon}=|\lambda_1-\lambda_2|$, where $\lambda_1$ and $\lambda_2$ are the eigenvalues of $(\epsilon+\epsilon^T)/2$, and $\epsilon^T$ is the transpose of $\epsilon$.  We view this field $\tilde{\epsilon}$ as a measure of the capacity of $\Vec{c}$ to induce shear strain at each location in the material.

We are also interested in quantifying correlations between this local shear strain and \dtwomin. To do so, we perform persistent homology clustering of the shear strain field, and then employ a modified normalized mutual information (NMI) metric \cite{stanifer_avalanche_2022} between the \dtwomin and local shear strain clusters, described in Appendix~\ref{app:clusters}.

\subsection{Randomly Oriented inclusions in Elastoplastic Models}
\label{random_epm}

Here we describe the EPM used to mimic AQRD at the mesoscopic level. 
Our version of the elasto-plastic model corresponds to a cellular automaton in a two-dimensional square lattice with periodic boundary conditions.  Originally, the model was conceived to represent incompressible, homogeneous, and isotropic materials under a simple shear deformation protocol. In this case however, the deformation is not a simple shear one, so we need to modify the model accordingly.  As in previous work~\cite{Rossi2022}, we focus on a single shear-stress component that we denote $\sigma$, which is obviously an approximation, even more in this setting. To each site $i$ of the square lattice of linear size $\mathcal{L}$ we assign a coarse-grained local stress value $\sigma_i$ in the presence of an external strain $\gamma$.  As such stress goes above a stability threshold $\sigma_{\rm th}$, the site yields and its stress drops by a quantity $\delta \sigma$ chosen from $P(\delta \sigma) = \frac{1}{\lambda}e^{-\delta \sigma/\lambda}$.
For this work we fix $\lambda = \sigma_{\rm th} = 1$.
The initial condition at $\gamma=0$,
$\sigma_{i}(\gamma=0)$, is independently and identically drawn from a probability distribution $P_0(\sigma_{i})$, given by
\begin{equation}
    P_0(\sigma_{i})= \frac{(1-\sigma_i^2)}{\mathcal{N}} e^{-\sigma_i^2/(2 R^2)}, \quad \sigma_i \in [-1,1],
\end{equation}
where $R$ is a parameter that characterizes the initial stability of the solid, which in real systems depends on the preparation protocol, and $\mathcal{N}$ is a normalization constant. 
The main difference with standard versions of the EPM studied in the past comes from the kernel used to redistribute the stress drop.
When the unstable site $i$ yields, its stress drop  contributes to other sites through a discrete stress propagator $G$ such that 
\begin{equation}\label{eq:sigmaprop}
    \sigma_j \rightarrow \sigma_j + G_{j,i} \delta \sigma_i ,
\end{equation}
for every $j$, where $\delta \sigma_i$ is the stress drop at site $i$ and $G_{i,i} = -1$.
A common choice for $G_{j,i}$ is the Eshelby elastic stress propagator, which in the case of a 2D continuous material is given by
\begin{equation}
G^E(\textbf{r}_{j,i}) = \frac{\cos(4 \theta_{j,i})}{\pi r_{j,i}^2},
\end{equation}
where $\mathbf{r}_{j,i}$ denotes the vector between sites $i$ and $j$, and $r_{j,i}$ and $\theta_{j,i}$ are, respectively, its magnitude and the angle it makes with the shearing direction.
Here we modify the propagator to mimic the AQRD protocol. 
When the correlation length is small, less than or equal to the typical size of a localized rearranging region, we can imagine that each site in the EPM will yield in a random direction, so that the Eshelby kernels are not aligned anymore along the direction of the shearing force.
In order to reproduce this we change the definition of the stress propagator to
\begin{equation}
G_i^{\rm RO}(\textbf{r}_{j,i}) = \frac{\cos(4 (\theta_{j,i} + \phi_i))}{\pi r_{j,i}^2},
\end{equation}
as in~\cite{Ozawa_2023}, which is now dependent on the site that yields and is rotated by an angle $\phi_i$ drawn from a uniform distribution in the range $[-\pi/4,\pi/4]$.
The generalization to correlated input fields is then obtained by choosing the $\phi_i$'s to be spatially correlated with a certain correlation length. 
Another difference with respect to the EPM in the AQS regime is that the update algorithm does not relax all the sites in parallel, but it relaxes at each step the least stable one.

\begin{figure*}
	\centerline{\includegraphics[scale=0.63]{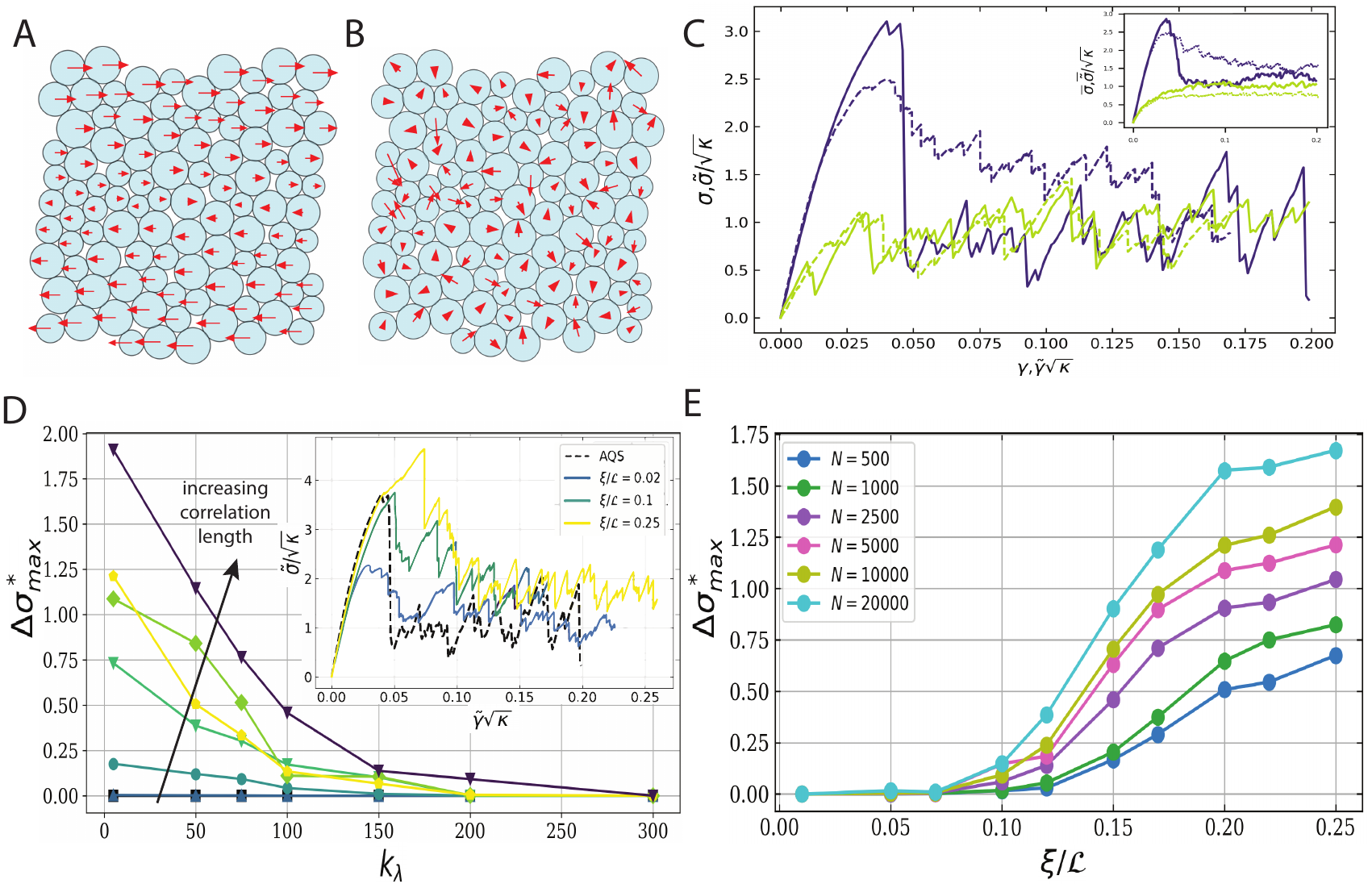}}
	\caption{\textbf{(A)} Input displacement field induced in linear response due to Lees-Edwards boundary conditions under AQS. \textbf{(B)} Example input displacements applied to particles with the smallest $\xi =1$ under AQRD. \textbf{(C)} Stress vs. strain curves for $k_\lambda=5$ (brittle) in purple and $k_\lambda=1000$ (ductile) in yellow for $N=5000$. Solid lines represent AQS and dotted lines represent AQRD (AQRD effective stress $\tilde{\sigma}/ \sqrt{\kappa}$ and strain $\tilde{\gamma} \sqrt{\kappa}$ given by Eq.~\ref{scaling}) with $\xi/\mathcal{L}=0.01$. Ultrastable systems ($k_\lambda=5$) show no large discontinuous stress drops under AQRD. Inset shows the same curves averaged over an ensemble of $20$ packings. \textbf{(D)}:~Average maximum stress drop $\Delta \sigma_{max}^*$ for a system size of $N= 5000$ vs the material preparation parameter $k_{\lambda}$ (lower  $k_{\lambda}$ is more ordered/stable) for different input field correlation lengths in AQRD \ag{($\xiol = 0.02, 0.05, 0.1, 0. 15, 0.2, 0.25$)  and AQS (highest purple curve marked with upside down triangles). Inset shows the stress strain curves for three different values of $\xiol$ and AQS (dotted line). The packing is ultrastable with $k_\lambda$ = 5 and system size $N = 5000$. The AQS curve shows the largest stress drop.} \textbf{(E)} Average maximum stress drop $\Delta\sigma_{max^*}$ for AQRD with $k_\lambda= 5$ versus input correlation length $\xiol$ for different system sizes.}
	\label{fig:1}
\end{figure*}

\section{Results}
\subsection{Yielding in active versus sheared granular solids}

We first investigate the stress-strain response in simulations as a function of the amount of disorder and the correlation length of the random `input' AQRD driving field relative to the box size, $\xiol$, and compare to the response of a sheared system under AQS. Fig.~\ref{fig:1}A is a schematic of the input displacements that are applied to each particle in AQS and Fig.~\ref{fig:1}B shows displacements in AQRD with the smallest resolvable correlation length $\xi=1$. Fig.~\ref{fig:1}C shows the stress-strain response under AQS (solid lines) and AQRD with $\xiol=0.01$ (dashed lines) for a system of $N=5000$ particles prepared with two different degrees of initial disorder -- $k_\lambda=5$, ultrastable, low disorder (dark purple) and $k_\lambda=1000$ low stability, high disorder (light green). 

As expected from previous work on systems prepared via swap Monte Carlo~\cite{Ozawa2018}, under AQS the ultrastable system is highly brittle, with large stress drops that correspond to system-spanning avalanches, while the low stability system is ductile with no large stress drops.  Just as in swap Monte Carlo systems, breathing particle systems show signatures consistent with a disorder-controlled transition similar to the one observed in the athermally-driven RFIM;
i.e., numerical data for different system sizes (Supplemental Fig.~\ref{fig:SI_aqsstats}) are consistent with the existence of a critical disorder at which avalanches become system spanning and the stress drops persist in the thermodynamic limit.

Previous work has also demonstrated that for strains below the macroscopic yielding transition, the statistical behavior of AQS and AQRD systems is identical up to the scale factor $\kappa$ described in Appendix~\ref{A:stress_def_aqrd}~\cite{morse_direct_2021}. In addition, infinite-dimensional mean field theory predicts that the dynamics should be the same~\cite{agoritsas_mean-field_2021} at all strains. However, we see in Fig.~\ref{fig:1}C that the yielding behavior of the ultrastable AQRD system (dashed purple line) deviates strongly from the AQS response (solid purple line) as the system approaches the yielding transition.  The ensemble averaged response (Fig.~\ref{fig:1}C inset) is also quite different.

To quantify this effect, we study the statistics of the maximum stress drop $\Delta \sigma_{max}$ observed along each stress-strain trajectory, as a function of the initial packing stability $k_{\lambda}$ for different types of input displacement protocols, Fig.~\ref{fig:1}D. To account for system size effects, we plot $\Delta \sigma_{max}^*=\langle\Delta \sigma_{max}\rangle - \langle\Delta \sigma_{max}\rangle_{\xi/\mathcal{L}=0.01,k_{\lambda}=300}$ where we have subtracted a baseline value for small avalanches $\langle \sigma_{max}\rangle_{\xi/\mathcal{L}=0.01,k_{\lambda}=300}$ for our system size of $N= 5000$, as has been done previously~\cite{Ozawa2018}. Different colors/symbols correspond to different correlations of the input driving field, ranging from AQS, with a $\xiol$ of 0.5 (half the box size, black boxes) to a $\xi/\mathcal{L}$ of $0.01$. Supplementary Fig.~\ref{fig:baseline} confirms that the baseline value for stress drop size $\langle\Delta \sigma_{max}\rangle_{\xi/\mathcal{L}=0.01,k_{\lambda}=300}$ becomes small as the system size $N$ increases.

Then, Fig.~\ref{fig:1}D shows that the largest stress drop is the same as this baseline for all values of the initial input disorder $k_{\lambda}$ when the input field is uncorrelated from particle to particle (blue), and the stress drop shows increasing departures from this baseline as the correlation length increases. \ag{The inset in Fig.~\ref{fig:1}D shows stress-strain curves for an ultrastable packing being driven by fields with varying correlation lengths and one curve for AQS. The discontinuous stress drops grow larger as the driving field becomes more correlated. The statistics for the magnitudes of the AQRD stress drops are shown in Fig.~\ref{fig:1}E.}   

We next numerically analyze whether this data is consistent with a phase transition as a function of the \emph{correlation length $\xi$}, instead of the disorder as in RFIM. Fig.~\ref{fig:1}E shows the magnitude of the average stress drop $\langle\Delta \sigma_{max}\rangle $  as a function of the input correlation length of the AQRD field $\xi/\mathcal{L}$, for different system sizes $N$. 

\mlm{
Our key numerical observation is that the magnitude of stress drops increases for values of $\xi/\mathcal{L}$ larger than about seven percent for all values of $N$. With the current data it is difficult to say whether this increase is a gradual crossover or a sharp phase transition that occurs at a specific value of $\xi/\mathcal{L}$, akin to the RFIM-like one previously observed as a function of the disorder.
}



\subsection{Yielding in randomized elastoplastic models}

\begin{figure*}
	\centerline{\includegraphics[scale=0.62]{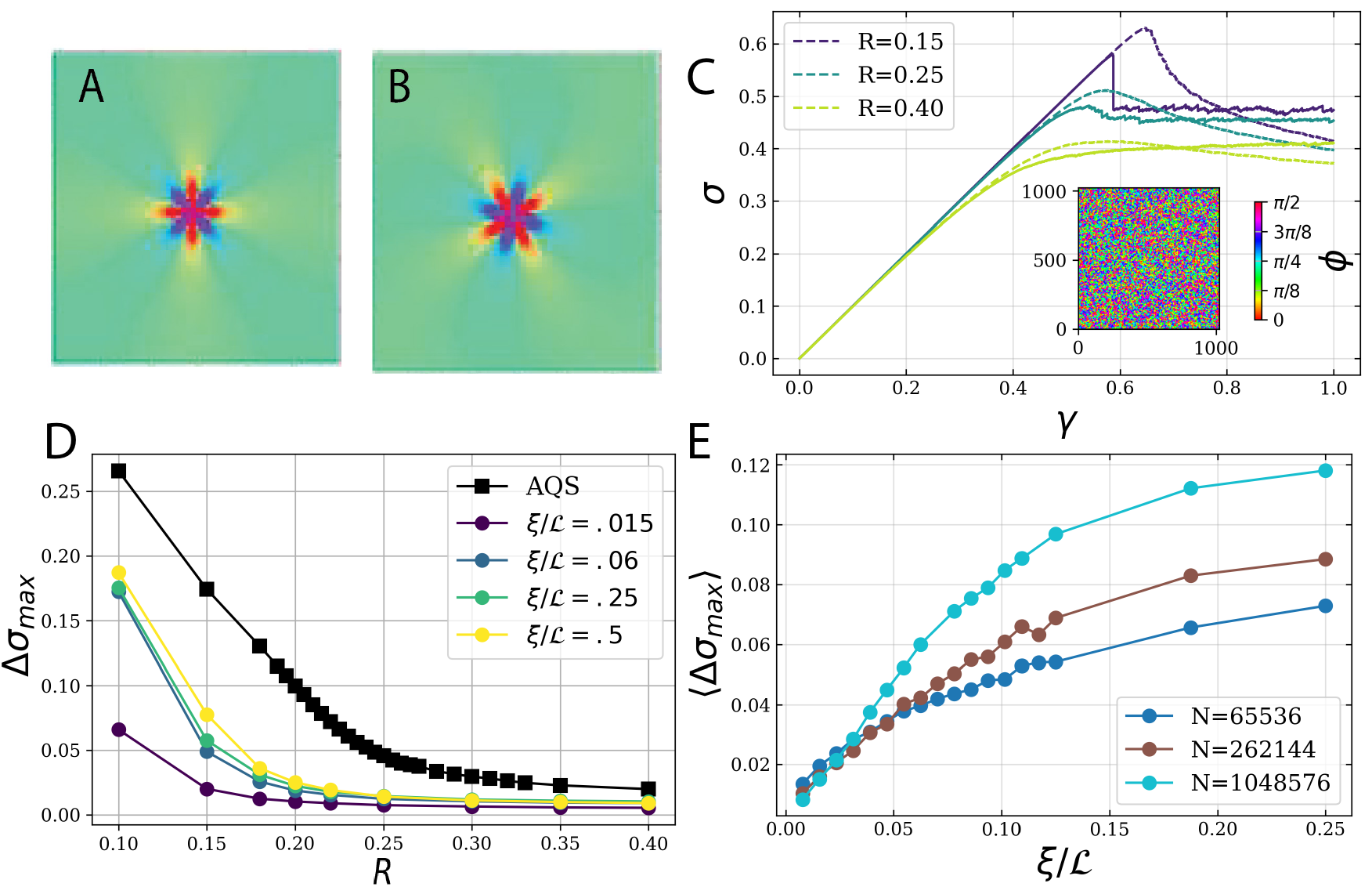}}
	\caption{\textbf{(A)} Standard orientation of Eshelby propagator for a yielding site \cite{Eshelby_1957, schall2007structural}. \textbf{(B)} Rotated Eshelby propagator for a yielding site in the modified EPM. \textbf{(C)} Stress vs. strain curves for three values of the EPM disorder $R$ in a system of $N=1048576$ ($\mathcal{L} = 1024$). Solid lines represent standard AQS EPM and dashed lines are for the EPM with randomly-oriented Eshelby kernels with $\xi = 8$ and $\xi/\mathcal{L}=0.007$. \textbf{(D)} Average maximum stress drop $\langle \Delta\sigma_{max}\rangle$ (averaged over initial stress realizations and orientational disorder) vs $R$ (low $R$ means less disorder/more stable).  \textbf{(E)} Average maximum stress drop $\Delta\sigma_{max}$ versus the correlation length of Eshelby orientations $\xiol$, for different values of the system size $N$ with disorder $R = 0.15$.  
	\label{fig:3}}
\end{figure*}

To make progress in understanding why ultrastable systems are no longer brittle under random forcing, we turn to EPMs, which have been effective in predicting the yielding behavior of granular systems under AQS~\cite{Ozawa2018, Rossi2022, malandro1998molecular, utz2000atomistic, utz2004athermal, schuh2003atomistic}. A key parameter in such EPMs is the disorder in the initial stress values of yielding sites: high disorder ($R$) corresponds to a broad distribution of initial stresses, whereas low disorder indicates a narrow one.  As has been shown previously~\cite{Rossi2022}, Fig.~\ref{fig:3}C illustrates that the standard EPM for AQS (solid lines) exhibits ductile behavior for high disorder $R$ and brittle behavior with a large stress drop for low $R$. This also qualitatively matches numerical particle simulation results for AQS (Fig.~\ref{fig:1}C).

As discussed in Section~\ref{random_epm}, in standard EPMs the rearrangements and corresponding Green's functions are oriented along the direction of the global applied shear. However, in AQRD we hypothesize these rearrangements are only correlated over the same length scale as the input applied field, $\xi$.  We therefore analyze the response of a randomized EPM, where Green's function orientations are correlated over a length $\xi$. Fig.~\ref{fig:3}A illustrates the Green's function associated with a standard EPM, where the directions of increased stress (red) lie parallel or perpendicular to the direction of applied shear. Fig.~\ref{fig:3}B illustrates the Green's function for a yielding site that is rotated with respect to the x- and y-axis, as appears in the modified EPM.

Next, we study the stress-strain curves for our modified EPM where the correlation is the smallest possible -- the orientation is uniformly random at every site. The results are shown by the dashed lines in Fig.~\ref{fig:3}C; the system no longer appears brittle, just as in the numerical simulations for AQRD.

To see if the modified EPM can also explain trends as a function of $R$ and $\xiol$, Fig.~\ref{fig:3}D shows the magnitude of the largest stress drop in the modified EPM as a function of $R$ and $\xiol$, in direct analogy to Fig.~\ref{fig:1}D.  We see good qualitative agreement between the prediction of the EPM and numerical simulations.  Similarly to Fig.~\ref{fig:1}E, Fig.~\ref{fig:3}E shows the scaling of the stress drop with $\xi$ as a function of system size, and again find a gradual onset of larger stress drops, this time with the onset starting from zero.

Comparing Fig.~\ref{fig:1}E and Fig.~\ref{fig:3}E, the shift in the onset of larger stress drops is perhaps not surprising. In the EPM, one can simulate much larger system sizes, but one cannot access the smaller correlation lengths that are accessible in a numerical simulation. This is because the smallest scale in an EPM is the grid spacing, which corresponds to the size of a mesoscopic region with a well-defined yield stress.  Since cores of localized rearrangements in 2D disk packings are typically estimated to be on the order of 20 to 50 particles~\cite{falk1998dynamics, stanifer_avalanche_2022}, AQRD input correlations lengths that are less than 5- to 7- particle diameters are not accessible in EPMs and Fig.~\ref{fig:3}E.




\subsection{Yielding angles are correlated over the same correlation length of the input field}

What mechanisms are driving these thermodynamically large stress drops in ultrastable systems, and what leads to their absence in low-correlation input AQRD systems? Previous work on AQS and slowly sheared systems has suggested that shear bands -- highly localized regions of non-affine plastic deformation oriented along the shear direction and percolating across the other dimensions -- enable the system to release macroscopic stresses in a single large avalanche~\cite{berthier2025yielding, Ozawa2018, fieldingdivoux2024ductile, fieldingbarlow2020ductile, singh2020brittle}. 

This is easy to explain in standard elastoplastic models for shear bands~\cite{Ozawa2018, Rossi2022}, where the Green's function falls off as \mlm{$1/r^{d}$}, and therefore the induced stress is highest at sites close to the original plastic event. Also, because the Green's function is non positive-definite, the stresses are positive along lines perpendicular and parallel to the shear, and negative along the diagonals (see Fig.~\ref{fig:3}A), resulting in a build-up of stresses along the parallel/perpendicular directions. Together, these two effects generate a shear band.

\begin{figure}
	\centerline{\includegraphics[scale=0.9]{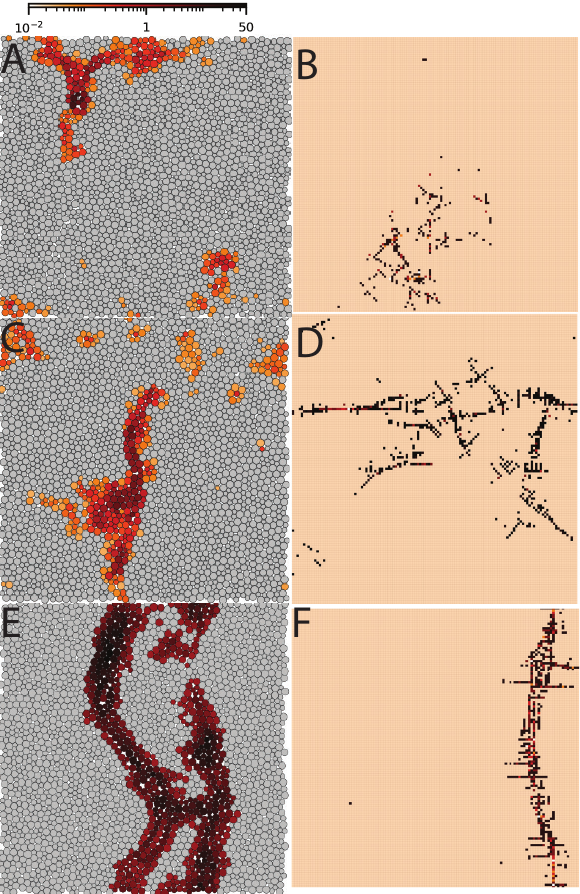}}
	\caption{\textbf{(A, C, E)} Intensity plots of \dtwomin for the largest stress drop in an example member of the ensemble for $\xiol= 0.05 ,0.15,0.25$ in a system of $N=2500$ and $k_\lambda = 5$. The darkest spots identify particles experiencing highly nonaffine rearrangements in a neighborhood of $5 \bar{r}$, as defined in Appendix~\ref{A:yieldingcorr}. \textbf{(B, D, F)} Spatial representation of modified EPM colored by number of yielding events per site for $\xiol = 0.023, 0.094, 0.25$ and $N=16384$ ($\mathcal{L}=128$).}  
	\label{fig:2}
\end{figure}

We observe an interesting change in the organization of shear bands as a function of $\xi/\mathcal{L}$ in both particle-based simulations (Fig.~\ref{fig:2}A-C-E) and in the modified EPM (Fig.~\ref{fig:2}B-D-F). At the largest values of $\xi/\mathcal{L}$ we observe features fairly similar to shear bands, while at smaller values of $\xi/\mathcal{L}$ we observe ``shear blobs'' that no longer span the system.


\begin{figure}
	\centerline{\includegraphics[scale=0.7]{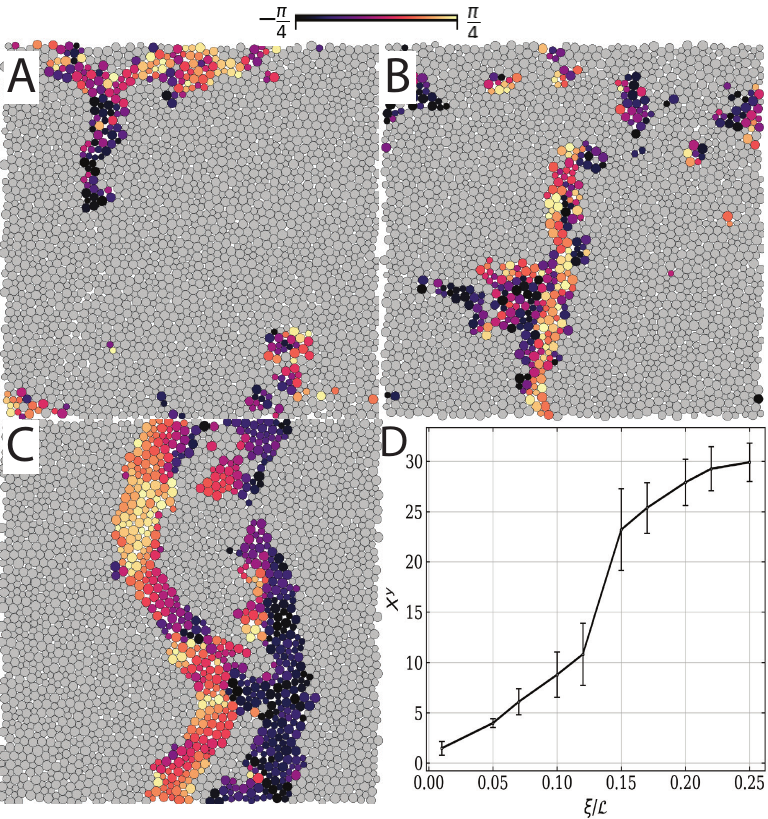}}
	\caption{\textbf{(A,B,C)} Intensity plots for the yielding angles $\theta^y$ of particles with high \dtwomin in the largest observed avalanches in AQRD under $\xiol =0.05, 0.15$ and  $0.25$ in a system with $k_\lambda =5$ and $N=2500$. \textbf{(D)} $X^y$ (defined in Appendix \ref{sec:Identifying_rearrangers}) in units of particle diameter versus $\xiol$ averaged over an ensemble of 20 systems and 80 stress drops for $N=2500$ ($L_y\approx L_x\approx50$).}
	\label{fig:4}
\end{figure}

Given the striking ability of the modified EPM to predict the numerical results, a natural hypothesis is that the orientation of the rearrangements within one of these shear blobs is correlated over the same correlation length, $\xi$, of the input field. 
This hypothesis is easy to test numerically in particle simulations. As described in Appendix~\ref{A:yieldingcorr}, we quantify the orientation of the plasticity in the neighborhood of each particle, and then compute the length scale over which this orientation is correlated, denoted $X^y$. As shown by the colorscale in Fig.~\ref{fig:4}A-C, the orientations vary rapidly when $\xi$ is small (panel A), and become correlated over the entire length of the shear blob at large $\xi$ (panel C).  A summary plot of $X^y$ vs. $\xi/L$ is shown in Fig.~\ref{fig:4}D.  \mlm{Consistent with this, the distribution of yield angles is nearly uniform at the lowest correlation lengths, but becomes sharply peaked (typically around one or two values) at the highest correlation lengths.} This represents a strong confirmation from our numerical simulations that the key assumption in the modified EPM -- that $\xi$ also defines the correlation length of orientation angles for Green's functions -- is correct.

\subsection{Spatial correlation between shear blobs and regions of high shear strain within the input field}

\begin{figure}
	\centerline{\includegraphics[scale=0.85]{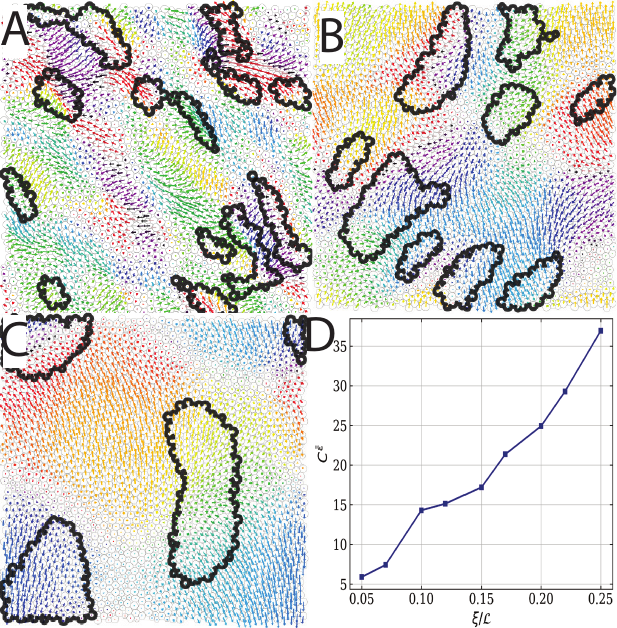}}
	\caption{\textbf{(A,B,C)} Example input displacement fields for AQRD with $\xiol = 0.05, 0.15, 0.25$ in a system of $N=2500$ particles. The length of the arrow indicates the magnitude of the displacement and the color indicates the orientation of each displacement vector. Clusters of high local shear strain identified as described in the main text are outlined in black. \textbf{(D)} Average length (major axis of best fit ellipsoid) of clusters of high local shear strain vs $\xiol$.}
	\label{fig:5}
\end{figure}

Another natural hypothesis is that the shear blobs occur at regions of high shear strain within the input AQRD field, and therefore the input field directly sets the lengthscale for the output \dtwomin field. 
To test this hypothesis, we first compute the local strain that arises from a displacement field on all the particles, and then use persistent homology to find discrete regions of highest local strain within each input field. Fig.~\ref{fig:5}A-C shows the input \mlm{displacement} field (arrows) and the identified local clusters (black outlines) for increasing $\xi$.  Fig.~\ref{fig:5}D validates our clustering algorithm: the discrete clusters of high local strain have the same length scale as the input displacement field.

We are now ready to directly quantify whether regions of high local input strain correlate with regions of high output plasticity. We use the proficiency described in Appendix~\ref{app:clusters} to perform this correlation. As discussed in Appendix~\ref{app:clusters}, the proficiency values are bimodal, allowing a clear separation between low values, corresponding to weak correlations between input and output clusters, and high values ($\chi > 0.06$), which indicate significant overlap between clusters.

\begin{figure*}
	\centerline{\includegraphics[scale=1.2]{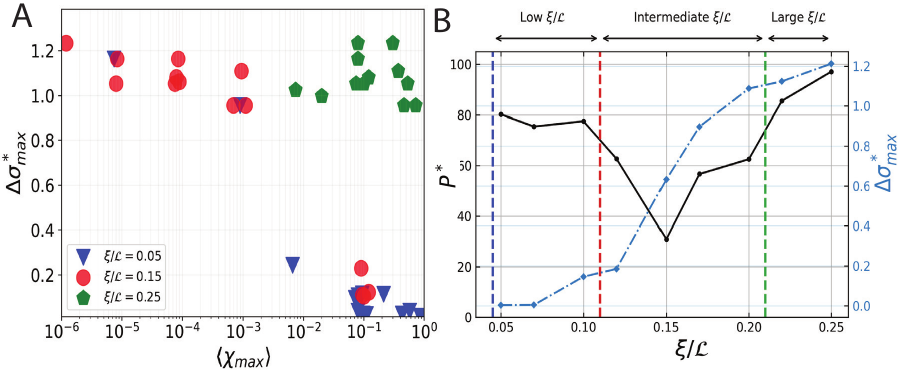}}
	\caption{\textbf{(A)} Scatter plot of the maximum stress drop $\Delta\sigma_{max}^*$ versus the average maximum proficiency $\langle \chi_{max} \rangle$ (for all clusters of \dtwomin in the largest stress drop) for three values of the input correlation strength $\xi/\mathcal{L}$, where for each $\xi/\mathcal{L}$ we show results for an ensemble of 12 ultrastable ($k_\lambda =5$, $N=5000$) initial conditions strained to 15\%. \textbf{(B)} Percentage of plastic events successfully predicted  by local shear strain ($P^*$, left axis), superimposed over the mean of $\Delta\sigma_{max}^*$ (right axis) for a system with $N=5000$ and $k_\lambda = 5$, versus $\xiol$.}
	\label{fig:6}
\end{figure*}

To visualize the relationship between stress drops and spatial organization of shear blobs, Fig.~\ref{fig:6}A is a scatter plot of the maximum stress drop vs. the average maximum proficiency (for all clusters of \dtwomin in the largest stress drop) for three values of the input correlation strength $\xi/\mathcal{L}$ (12 different ultrastable packings at each $\xi/\mathcal{L}$ ).  A first observation is that the different $\xi/\mathcal{L}$'s are clustered in three different regions of this phase space. 

\begin{itemize}
\item 
At low $\xi/\mathcal{L}$ (blue), the proficiency is typically high, meaning that the locations of high input strain precisely correlate with the output shear blobs. The maximum stress drop is very small, meaning the output shear blobs are small, and the regions of high local input strain are also small. So in this region of space, our hypothesis was correct -- the input field governs the output field.

\item
At very large values of $\xi/\mathcal{L}$ (green), the proficiency is also very high, but now the stress drops are quite large. Macroscopic stress drops correspond to system-spanning avalanches, and those shear bands happen in precisely the locations that one would predict from the input strain field. Again, the input field governs the output field.

\item
Most interesting are the results at intermediate $\xi/\mathcal{L}$ (red). In this case, the proficiencies are low, suggesting that the input strain is {\it not} a good predictor of the output plasticity. There were previous hints that something different is happening in this regime; as one can see in Fig.~\ref{fig:4}B and Fig.~\ref{fig:5}B, at $\xi/\mathcal{L} = 0.15$ there seems to be a mismatch between the size of the \dtwomin cluster, which appears nearly system-spanning, and the high-strain clusters, which are smaller.  Fig.~\ref{fig:6}A confirms that the avalanches are indeed system-spanning; the red data points do typically exhibit macroscopically large stress drops.
\end{itemize}

We speculate there must be an initially high-strain region that triggers an initial rearrangement, but because the system is inherently ultrastable, it is susceptible to an avalanche that turns into a shear band. Therefore the initial high strain region does not predict the full shear band, and the proficiency is low. In this case, the output field is self-organized, and is not directly governed by the input field.

To quantify this, we define $P^*$ as the percentage of plastic events that are successfully predicted solely by the input strain field, which is the number of stress drops with exactly one proficiency greater than $0.06$, and plot $P^*$ as a function of $\xiol$. This is shown as the left axis of Fig.~\ref{fig:6}B, where we highlight three regimes: low and high $\xiol$ where the input field largely predicts output rearrangements and plasticity, and intermediate $0.11< \xiol <0.21$ where the input field does not accurately predict the output field. The right axis of Fig.~\ref{fig:6}B shows the average maximum stress drop $\Delta \sigma_{max}^*$ as a function of $\xiol$, illustrating that the largest gradient of the curve also occurs in this intermediate regime.


\section{Conclusions and Discussion}
Our results demonstrate that, in contrast to predictions from infinite-dimensional mean field theory, the yielding behavior of dense active matter fundamentally differs from that observed under conventional shear deformation. Materials that are prepared to be highly stable/ordered, which would fail in a brittle manner under shear, flow in a ductile way under an applied random field.
This highlights that stability is not synonymous with brittleness and that the statistics, and specifically the correlation length, of the input driving field play a key role in determining emergent plasticity and yielding. This work shows that in highly stable systems there is an apparently continuous transition from brittle failure -- when the input field has a large correlation length -- to ductile flow when the field is uncorrelated.

This insight has allowed us to demonstrate that an EPM with a simple modification that accounts for these input correlation lengths provides a qualitatively predictive constitutive theory for dense active matter in the limit of slow driving. We have directly confirmed in simulations the key assumptions of this EPM -- that the orientation of rearrangements is correlated over the length of the input field $\xi$. This new EPM will provide a starting point for future work to design and control dense active matter systems.

In addition, it is possible to go beyond qualitative predictions in some cases. There are large regions of parameter space where the input field alone completely predicts the output plasticity.
This is quite different from sheared passive systems, where there is no information contained in the input field, because the strain is uniform everywhere. In such systems, immense time and effort has been spent on identifying soft spots or structural defects~\cite{manning2011vibrational, gartner2016nonlinear, richard_predicting_2020, zhang2020interplay} to predict the location and statistics of plastic flow. \mlm{Identifying structural defects in active systems can be challenging, as the disordered active field has a strong impact on plasticity that can invalidate standard methods~\cite{giannini2022searching}, though recent work has suggested useful extensions~\cite{hain2026}.}

We have demonstrated that for high and low values of the input correlation strength $\xi/\mathcal{L}$, one can predict the output plasticity with 80\% accuracy just from the input field. This  suggests that with proper design, flow in granular systems can be surprisingly highly controllable by an input active field. While the annealed structural disorder helps to dictate features of the macroscopic stress-strain curve, it is less important for predicting individual rearrangements.

The regimes where the input field does not dictate the output may be more interesting from a fundamental physics perspective. In these cases, there is an interplay between the annealed structural disorder and the disorder in the input driving field that leads to a self-organization of system-spanning shear bands.
This exciting regime will thus provide a difficult and strong test for quantitatively predictive constitutive laws.  Many different theories lead to constitutive laws that generate quite similar predictions for stress-strain curves and even spatial organization of avalanches, and it has historically been difficult to identify simulations or experiments that distinguish between them.  However, even our modified EPM fails to predict plasticity in this regime.

We expect that what is missing is the coupling between the structural disorder and the orientations of the Green's functions.  One simple way to encode this would be to include some correlations between energy barrier heights and Green's function orientations in the modified EPM. \mlm{One could additionally include explicit spatial correlations between energy barriers and orientations as observed in simulations.} A more difficult, but perhaps more realistic approach are so-called ``structural Elasto-plastic'' (stEP) models~\cite{zhang2020interplay, zhang2022structuro}, where researchers have shown that the structure and statistics of the annealed disorder (e.g. the energy barrier heights) are directly impacted by plasticity, in addition to the indirect coupling through the stress via the Green's function. \mlm{Identifying the locations and statistics of structural defects in active systems~\cite{hain2026} would help constrain these models.} Such explicit correlations between annealed structural disorder and input field disorder could significantly enhance our ability to predict flow patterns in this regime.

In addition, it seems likely that there is a direct coupling between the orientation of the strain from the input field and the inherent orientation of defects. Work on the yielding angle~\cite{desmarchelier2024topological, Patinet2016, xu2021atomic, rottler2018} suggests that structural defects do have preferred orientations, and so the input field could either reinforce or oppose the natural yielding direction, which would be another method to implement more precise control. \mlm{It should be possible to incorporate recent work to include active forces in computing energy barriers for defects~\cite{hain2026}, to account for these effects in more sophisticated EPMs.} \mlm{We also observed, in some cases, that individual localized plastic events in these active matter systems did not appear to be well-fit by an Eshelby kernel, which is why we chose a different method for quantifying the yielding angle compared to previous work~\cite{rottler2018}. In the future, it would be interesting to understand how the interactions between the input driving field and the local inherent orientation of the defect affect the intermediate-scale features of the resulting displacement fields.}


From a practical perspective, another important area for future study is the effect of finite persistence times, finite driving rates, and finite temperatures on flow patterns in dense active matter. Just as AQS has proven to be a very useful limit for considering sheared systems, our focus on the AQRD system in the same limit -- zero temperature, infinitely slow driving, infinite persistence time -- has allowed us to make useful progress on understanding fundamental principles. In sheared systems, EPMs have been extended to finite driving rates by accounting for effects like inertia and interrupted avalanches~\cite{nicolas_deformation_2018, liu2016driving}, \mlm{and recent work has expanded these ideas to dense active matter~\cite{mandal2020multiple, mandal2021study}}. We expect that a slowly time-varying input field corresponding to finite persistence times for active forces would generate similar perturbations, and it would be exciting to perform detailed comparisons between simulations and extensions of finite strain-rate EPMs as a function of the driving rate. It would be particularly interesting to know how the controllability by the input field changes as the input persistence timescale decreases.

Our initial work here paves a concrete path towards understanding the mechanisms that dense animal/bacterial collectives in nature could harness to drive functional emergent behavior.  It also highlights that the detailed dynamics of dense granular solids, long assumed to be difficult and perhaps intractable to predict, may be surprisingly controllable by an active input field. This leads to new ideas for design principles in dense amorphous matter that might be programmed to solidify and/or flow on command to perform functional tasks.

\section{Data availability}
The data used to generate all figures in the manuscript are available from Ref.~\cite{ghaznavi2025yielding}. 

\acknowledgments

We thank Gilles Tarjus, Julia Giannini, and Tyler Hain for fruitful discussions and Misaki Ozawa for his helpful insights regarding the code.
This research has been supported by NSF DMR-1951921 and NSF DMR-2532170, first FIS (Italian Science Fund) 2021 funding scheme (FIS783 - SMaC - Statistical Mechanics and Complexity) from MUR, Italian Ministry of University and Research and from the PRIN funding scheme (2022LMHTET - Complexity, disorder and fluctuations: spin glass physics and beyond) from MUR, Italian Ministry of University and Research.

\appendix

\section{Definitions and protocols for simulating AQRD}
\label{A:stress_def_aqrd}
A key feature of the input displacement field $\Vec{c}$ is the lengthscale over which the displacements are correlated.  We create an ensemble of input displacement fields by first defining a characteristic lengthscale relative to the size of the box $\xiol$, where $\xi$ is the correlation length and $\mathcal{L}$ is the box length in units of the particle diameter. Next, we generate a 2D vector-valued continuous gaussian random field (correlated over that lengthscale) defined over the two-dimensional domain, which also respects the periodic boundary conditions, as described previously~\cite{morse_direct_2021}. Then, the value of the displacement on each particle is given by the 2D vector defined by the Gaussian field evaluated at the particle center. Due to the periodic boundary conditions, the maximum correlation length possible without introducing artifacts that align the field with the box edges is $\mathcal{L}/4$. In addition, because we always coarse-grain the field to lie on particle centers, we cannot resolve correlation lengths below $1/\mathcal{L}$ for a given system size $N = \mathcal{L} \times \mathcal{L}$. In order to compare correlation lengths $\xiol$ over a broad range of system sizes $N$, we report correlation lengths in units of $\xiol$ and not~$\xi$.   

For completeness, the next few paragraphs summarize results from Ref.~\cite{morse_direct_2021}, which develops a method to compare the magnitude of input displacements $\Vec{c}$ from AQRD to the strain applied via Lees-Edwards boundary conditions in AQS.

 We note that in linear response, shearing the Lees-Edwards boundaries a distance $\gamma L_y$ is equivalent to applying a displacement to each particle of size $\gamma (y_i-L_{y}/2)$, where $y_i$ is the $y$-coordinate of the particle and $L_y$ is the length of the box in the $y$-direction. In a square box, this gives a total displacement magnitude of $\gamma L_{y} \sqrt{N/12}$, and we write that the effective strain $\Tilde{\gamma}$ due to a displacement $u= \gamma L_{y} \sqrt{\frac{N}{12}}$, where $L_y$ is the length of our box in the $y$-direction.  In our simulation we are applying displacements along a normalized vector field $\Vec{c}$. Our displacements, using our definition of the random strain, can be written as

\be\label{dx-i}
dx_i = du_i c_i =  d\gamma L_{y} \sqrt{\frac{N}{12}} c_i
\ee

We put tildes over the $\gamma$ and $u$ to indicate when they describe variables in a random field. For AQRD the relation becomes

\be\label{tilde}
\Tilde{u}_i c_i = \Tilde{\gamma} L_{y} \sqrt{\frac{N}{12}} c_i
\ee
The standard definition of the stress is $\frac{dU}{d\gamma}$. To find our random stress in AQRD we need to replace the $\gamma$ in the expression with $\Tilde{\gamma}$. So, we need to find 

\be\label{rand_stress}
\Tilde{\sigma} = \frac{dU}{d\Tilde{\gamma}} = \Sigma(\frac{\p U}{\p x_{i}^{||}}\frac{dx_{i}^{||}}{d\Tilde{\gamma}}+\frac{\p U}{\p x_{i}^{\perp}}\frac{dx_{i}^{\perp}}{d\Tilde{\gamma}})
\ee

Since we have performed a constrained minimization on our system and subtracted off any forces in the direction of $c$, the only remaining force on the particles must be parallel to $c$, which leaves us with only the first term in the above expression. So we have

\be\label{stress}
\Tilde{\sigma} =  \Sigma(\frac{\p U}{\p x_{i}^{||}}\frac{dx_{i}^{||}}{d\Tilde{\gamma}})
\ee

There are two derivatives in the expression above. The second one can be evaluated using our definition of the random strain and is equal to $c_{i}L_{y}\sqrt{\frac{N}{12}}$. The first is the component of the force parallel to $c$ and as such is just the projection ${F}_{c}$, giving us the expression for the random stress

\be
\label{stress_f}
\Tilde{\sigma} = \braket{F}{c}L_{y}\sqrt{\frac{N}{12}}
\ee

An infinite-dimensional mean-field calculation \cite{agoritsas_mean-field_2021, rainone_following_2016, biroli_breakdown_2016, biroli_liu-nagel_2018, urbani_shear_2017, altieri_mean-field_2019} suggests that AQRD and AQS should have the same dynamics up to a scale factor:

\be\label{mft}
\gamma_{AQS}=\Tilde{\gamma}\frac{\sqrt{\mathfrak{F}}}{l} \ , \qquad \sigma_{AQS}=\Tilde{\sigma}\frac{l}{\sqrt{\mathfrak{F}}}
\ee

Here, $\mathfrak{F}$ is the variance of the strain between neighbors in the input field and $l$ is the typical distance between particles. Clearly the variance in strain across neighbors decreases as $\xi$ increases, and AQS has no variance in the strain. Since in linear response the shear modulus is just the ratio, $\mu = \sigma/\gamma$, the theory predicts that the ratio between the  two shear moduli $ \mu_{AQRD} / \mu_{AQS}$ should be given by $\frac{\mathfrak{F}}{l^2}$. In other words, a material has a stiffer response to an input field that induces larger local strains.

Previous numerical work indicates that this mean-field prediction gives the right trends but does not precisely hold in $d=2$~\cite{morse_direct_2021}. Instead, the authors empirically \emph{measured} the modulus associated with elastic (monotonically increasing) branches of the stress-strain curves in each simulation and defined an empirical scalar $\kappa \equiv \mu_{AQRD} / \mu_{AQS}$ for each value of $\xi$. They showed the mean-field-inspired rescaling,
\be
    \label{scaling}
    \sigma=\frac{\Tilde{\sigma}}{\sqrt{\kappa}}, 
    \qquad \gamma=\Tilde{\gamma}{\sqrt{\kappa}},
\ee
precisely collapses stress-strain curves and avalanche statistics in the pre-yielding regime~\cite{morse_direct_2021}. Therefore, in our work here we compute $\kappa \equiv \mu_{AQRD} / \mu_{AQS}$ numerically from our ensemble of simulations and use it rescale all of our stresses and strains. This ensures that the trivial rescaling of the stiffness due to $\xi$ is removed, and ensures all stress-strain curves are statistically identical up to the yielding point.

Previous work has identified a step size of $10^{-4}$ as a sufficiently small step size in AQRD to reliably generate system dynamics~\cite{morse_direct_2021}. To accurately compare AQRD and AQS, we must next compute the scaling factor $\kappa$ between the two described in the previous section. We start off by simulating steps of size 0.0001 in AQRD until the response of the system is no longer elastic to first order, so we can get an accurate estimate for the random shear modulus $\tilde{\mu}_0$. This step size represents $\Tilde{u}$ i.e. the distance travelled in the $c$ direction. To get the average strain for these steps we use $\Tilde{u} = \Tilde{\gamma} L_{y} \sqrt{\frac{N}{12}}$. This gives us 

\be
    \Tilde{\gamma} = \frac{\Tilde{u}}{L_y\sqrt{\frac{N}{12}}}
\ee

We then calculate the random stress at each strain step before the response stopped being elastic. At each step we also calculate $\tilde{\mu}_i$ and we take the average $\langle \tilde{\mu}_0 \rangle$. We similarly get a value for the average AQS shear modulus $\langle \mu_0 \rangle$. 
Using this we find $\kappa = \frac{\langle \tilde{\mu}_0 \rangle}{\langle \mu_0 \rangle}$, and compute the total number of steps required in AQRD to match 20\% strain in AQS.

\section{Identifying rearranging particles}

\label{sec:Identifying_rearrangers}
Plasticity in disordered systems is well captured by $D^2_{min}$, a measure of the nonaffine motion~\cite{falk1998dynamics}. $D^2_{min}$ compares two configurations of a system over a specified radius, in this case five average particle radii:

\begin{equation}
\label{eq:d2}
D^2_{min,i}\left(\overrightarrow{X}_1,\overrightarrow{X}_2\right)=\sum_{j:r_{ij}<5\bar{r}}\left(\overrightarrow{r_{ij}}_2- \epsilon_i \overrightarrow{r_{ij}}_1\right)^2,
\end{equation}
where $\overrightarrow{X}_1$ and $\overrightarrow{X}_2$ represent the two configurations being compared, $r_{ij}$ is the distance between particles $i$ and $j$, $\bar{r}$ is the average particle radius, $\overrightarrow{r_{ij}}_1$ and $\overrightarrow{r_{ij}}_2$ are the vectors that separate particles $i$ and $j$ in the first and second configuration respectively, and $\epsilon_i$ is the best-fit affine transformation that minimizes $D^2_{min,i}$.

As is standard~\cite{falk1998dynamics}, to compute $D^2_{min}$ one must choose a lengthscale for the neighborhood over which the affine and non-affine transformations are computed.  Consistent with previous work, we choose $5\bar{r}$. Smaller lengthscales may cause an error in the algorithm because some neighboring particles will not be included in a neighborhood which is too small, while larger lengthscales result in $D^2_{min}$ fields that are more homogeneous and fail to capture localized rearrangements. 

\section{Calculation of yielding angle and correlation length}
\label{A:yieldingcorr}
In computing the nonaffine displacement field $D^2_{min}$~\cite{falk1998dynamics}, one must also compute the best-fit affine transformation tensor $\epsilon$ in a neighborhood around the particle of interest. A typical choice of this neighborhood is $5 \bar{r}$, where $\bar{r}$ is the mean radius of particles in the packing.

Here, we are interested in using the \emph{affine} part $\epsilon$ to characterize the shear strain in the random input fields. Specifically, the local deviatoric shear strain tensor in the region is given by the traceless symmetric component of~$\epsilon$:
\be
\mathcal{M}^* = \mathcal{M}_{shear  strain} = \frac{1}{2}[\mathcal{M}+\mathcal{M}^T-\frac{Tr(\mathcal{M})}{2}\mathbf{I_n}]
\ee
We choose the eigenvector of this tensor $\Vec{\lambda}_{\mathcal{M}^*}$ with the largest associated eigenvalue and compute the angle it makes with the $x$-axis. We define this angle as the yielding angle $\theta^y$. \ag{This definition is similar to previous work~\cite{rottler2018} using an Eshelby kernel to fit for the strain; key differences include that our reference axis is the $x$-axis, whereas they used the principal direction of the macroscopic shear, and that we choose not to explicitly fit the Eshelby form because the disordered input field makes the plastic output field noisy and difficult to fit.}

To quantify the characteristic length scale over which local yielding angles remain correlated, we compute the spatial autocorrelation function. Specifically, each particle $i$ has a position $\mathbf{r}_i$ and a scalar yielding angle $\theta_i$. We first subtract the mean $\bar{\theta}$ to obtain zero-mean values $\delta \theta_i = \theta_i - \bar{\theta}$ and define the two-point autocorrelation as
\begin{equation}
C(\mathbf{r}) = \left\langle \delta \theta_i \, \delta \theta_j \right\rangle_{|\mathbf{r}_i - \mathbf{r}_j| \in [r,\,r+\Delta r]},
\end{equation}
where the average is taken over all particle pairs $(i,j)$ separated by a distance within the bin $[r, r+\Delta r]$.  
The corresponding normalized autocorrelation function is
\begin{equation}
\rho(r) = \frac{C(r)}{C(0)},
\end{equation}
where $C(0)$ is the variance of the yielding angle field.  
We identify all particle pairs within a cutoff distance $r_{\max}$ and compute $\rho(r)$ in evenly spaced radial bins.  

The resulting correlation function $\rho(r)$ typically decays monotonically with distance.  
We extract a characteristic correlation length $X^y$, which we call the yielding length, by fitting the short-range decay to an exponential:
\begin{equation}
\rho(r) \approx A \, e^{-r/X^y}+c.
\end{equation}

\section{Methods for quantifying clusters and overlaps between clusters}
\label{app:clusters}

We briefly review our adaptation of persistent homology techniques~\cite{otter2017roadmap_persistenthomology} for identifying clusters in disordered fields. The algorithm to generate a persistence tree begins by identifying the value of the field on each particle and sorting the list from highest to lowest value. Starting with the highest value, the algorithm iterates through this list, and for each particle finds neighboring particles that are close to it, defined using Lees-Edwards boundary conditions. If the particle does not have any neighboring clusters (groups of particles), then it represents a local peak and a new cluster is started just for that particle. If the particle has only one neighboring cluster, it is added to that cluster. If the particle has multiple neighboring clusters, it merges the parent clusters and starts a new child cluster that includes the current particle. The birth value of a cluster is the \dtwomin value of the particle  that started it. The death value is the \dtwomin value of the particle that ended it (caused it to merge with another). The persistence value for a cluster is given by $birth-death$. 

Once the persistence tree is built with all clusters, we wish to prune clusters that are too small or not significant. If a cluster has too few or too many particles -- here we chose $15$  and $\frac{3}{4} N$ particles, respectively, as these thresholds -- it is considered not significant and pruned (removed). The algorithm then iterates through clusters in reverse order, starting from the last one discovered. If a cluster has two or more parent clusters with a higher persistence it is also pruned. If a cluster is kept, its parent clusters are pruned to avoid redundancy. After the pruning process, we are left with a set of clusters with high persistence that represent meaningful patterns in $D^2_{min}$.

Once clusters are identified, we often wish to determine whether a cluster from one field is predictive of a cluster appearing in the same place in a different field.  To do so, we compute a normalized mutual information metric, which was first described in~\cite{stanifer_avalanche_2022}. 

The standard mutual information between two datasets \(I\) and \(J\) is defined as $M(I,J)=\sum_{x\in [I]} \sum_{y\in [J]} p_{x,y} \log_{2}\left( \frac{p_{x,y}}{p_x p_y}\right),$ where \(p_x\) and \(p_y\) denote the probabilities of a particle belonging to set \(x\) or \(y\), respectively, and \(p_{x,y}\) is the joint probability of a particle belonging to both. Probabilities are computed as the fraction of particles within a given set relative to the total number of particles observed in the simulation window.  

To distinguish between correlations arising from direct overlap of two fields and those originating from overlap with the complement of a field, we adopt a modified definition of the mutual information:
\begin{align}
&\tilde{M}(I,J)= \notag \\
&\sum_{x\in [I,\sim I]} \sum_{y\in [J,\sim J]} p_{x,y} \log_{2}\left( \frac{p_{x,y}}{p_x p_y}\right) \,\text{sgn}\left(p_{I,J}-p_I p_J\right).
\end{align}
The inclusion of the sign function ensures that correlations are counted as positive when the overlap between \(I\) and \(J\) exceeds the statistical expectation, and negative when \(I\) is more strongly correlated with the complement of \(J\). This adjustment prevents high values of mutual information that can otherwise occur due to overlap between one set and the complement of another.  The information entropy of a field, \(\tilde{H}(I)\), is obtained as the self-information, $\tilde{H}(I)=\tilde{M}(I,I).$

Finally, we report a normalized version of this metric, which we term the \emph{proficiency}: $\mathcal{\chi}_{IJ}=\frac{\tilde{M}(I,J)}{\tilde{H}(I)}$.  This quantity measures how well the regions of high plasticity are predicted by clusters of high local shear strain.

\section{Supplementary data analyzing avalanches, clustering, and cluster overlaps}

Fig.~\ref{fig:1}C in the main text showed stress-strain curves for packings with varying disorder being driven by an uncorrelated random field.

\begin{figure}[t]
	\includegraphics[scale=0.61]{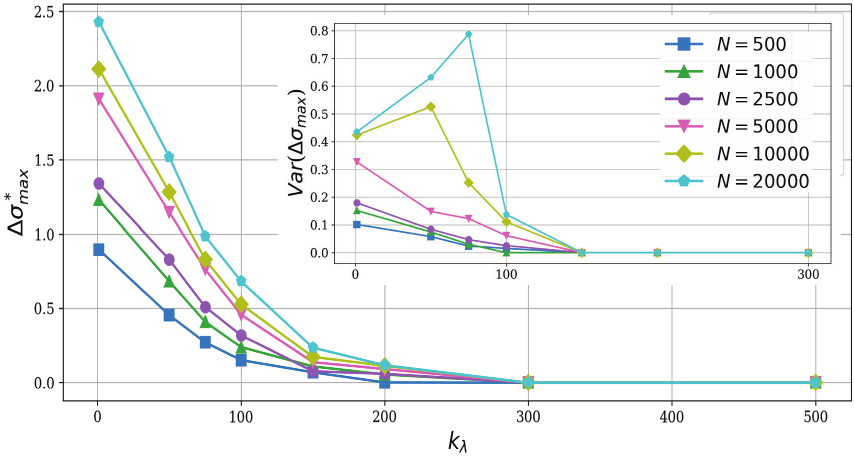}
	\caption{Mean ($\Delta \sigma_{max}^*=\langle\Delta \sigma_{max}\rangle - \langle\Delta \sigma_{max}\rangle_{k_{\lambda}=300}$) and variance (inset) of $\Delta\sigma_{max}$ vs $k_{\lambda}$ (low $k_\lambda$ corresponds to low initial disorder) for increasing system sizes under AQS.}
	\label{fig:SI_aqsstats}
\end{figure}

Previous work has carefully characterized the effect of material preparation on the largest stress drop in AQS, using swap Monte Carlo methods to prepare ultrastable packings~\cite{Ozawa2018}. In Fig.~\ref{fig:SI_aqsstats} we demonstrate that similar results are generated using the breathing particle method to prepare ultrastable packings.

\begin{figure}[t]
	\centerline{\includegraphics[scale=0.66]{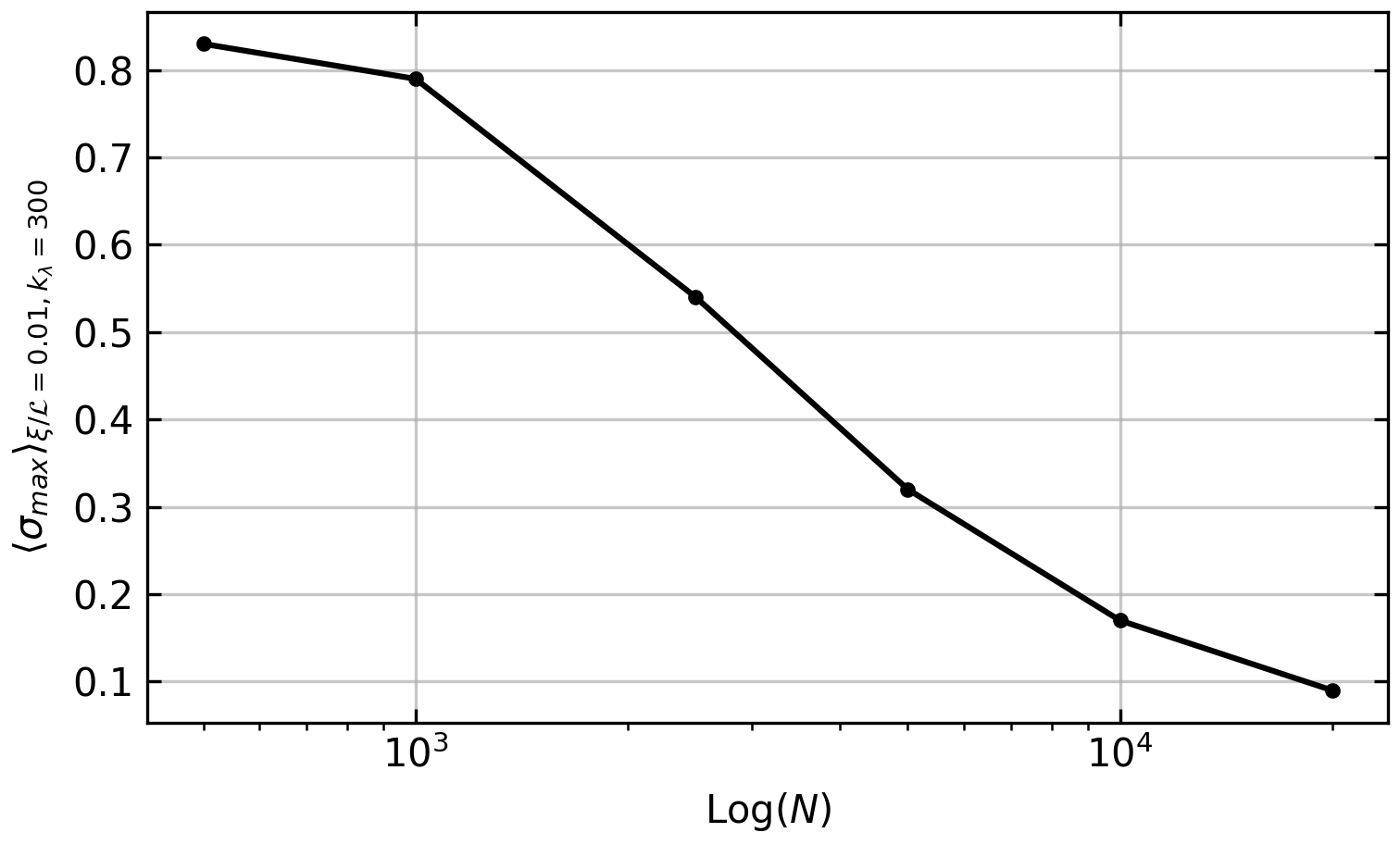}}
	\caption{Values of the baseline $\langle \sigma_{max}\rangle_{\xi/\mathcal{L}=0.01,k_{\lambda}=300}$, which been subtracted off in the AQRD stress drop values reported in the main text, as a function of the system size $N$.  For each system size, this represents the characteristic size of a 'small', non-system spannning avalanche for a ductile system driven by an uncorrelated input field.}
	\label{fig:baseline}
\end{figure}

In addition, we characterize the small avalanches that occur in ductile systems under driving by AQRD in Fig.~\ref{fig:baseline}. As expected, the size of small avalanches decreases as a function of system size $N$. As in previous work~\cite{Ozawa2018}, for each system size we subtract off this baseline average value from the largest stress drop when computing $\Delta \sigma_{max}^*$ in the main text.

\begin{figure}[H]
	\centerline{\includegraphics[scale=0.18]{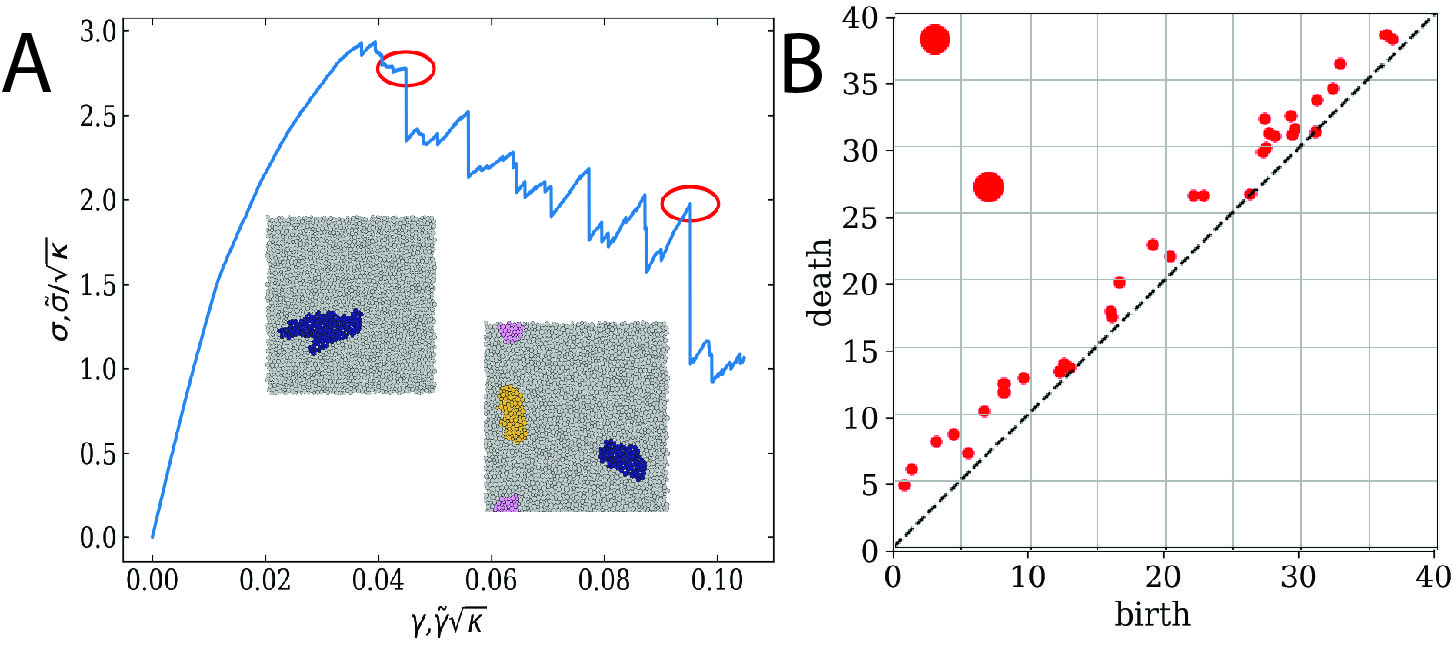}}
	\caption{\textbf{(A)} \dtwomin clusters for the two largest (circled in red) stress drops for AQRD in  a system of $2500$ particles ($k_\lambda=5$) at $\xiol =0.10$. \textbf{(B)} Birth and death plot for cluster formation in persistent homology. The two points far from the $x=y$ line have high persistence (were formed at a high \dtwomin value and ended at a very low one). They represent significant local peaks in \dtwomin while the rest represent noise.}
	\label{fig:SI_birthdeath}
\end{figure}

Fig.~\ref{fig:SI_birthdeath} highlights some details about our clustering procedure, persistent homology. Panel A shows the observed clusters at the two largest stress drops when an ultrastable system is driven by field with $\xiol = 0.10$. Panel B shows a birth death plot for the clustering algorithm. As mentioned in Sec \ref{sec:analysis_simulation}, the persistence of a cluster is defined as $birth-death$. Looking at panel B, we can conclude that the distribution of the persistence for all clusters (red dots) is bimodal, with a peak at low persistence due to noise; and two data points with a substantially higher value of the persistence (larger red dots).

\begin{figure*}
	\centerline{\includegraphics[scale=1.14]{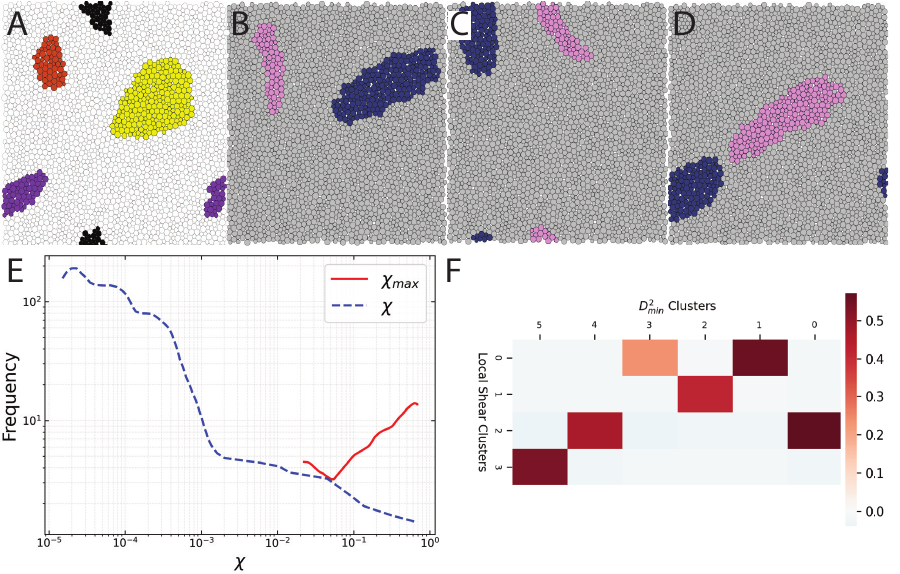}}
	\caption{ \textbf{Illustration of computing the fraction of rearrangements predicted by the input field for $\xiol =0.10$. (A)} Input shear clusters for a sample input field with $\xiol= 0.10$. \textbf{(B,C,D)} \dtwomin\, clusters for the three largest stress drops for an ultrastable packing driven by the input field in (A). \textbf{(E)} Distribution of proficiency values for all local shear clusters with all \dtwomin\, clusters. Blue dotted line shows $\chi$ for all pairs of clusters except $\chi_{max}$ for each \dtwomin\, cluster. Red line shows the distribution of $\chi_{max}$ for all \dtwomin\, clusters. \textbf{(F)} Heatmap showing the proficiency values between all pairs of clusters seen in \textbf{(A)} with \textbf{(B,C,D)}. The predicted fraction is the number of \dtwomin\, clusters that have a corresponding input shear cluster with a proficiency greater than $0.06$.}
	\label{fig:SI_fig6(a-f)}
\end{figure*}

Fig.~\ref{fig:SI_fig6(a-f)}A shows the local shear strain clusters in an AQRD field with $\xiol = 0.10$. Fig.~\ref{fig:SI_fig6(a-f)}B-D show the \dtwomin \, clusters at the largest stress drops. The heatmap in Fig.~\ref{fig:SI_fig6(a-f)}F shows that each \dtwomin \, cluster has a high overlap with at least one cluster of high local shear strain. We see the same information for all possible cluster pairs in Fig.~\ref{fig:SI_fig6(a-f)}E for $200$ avalanches. The red curve shows a clear bimodal distribution of the maximum proficiency values, allowing us to choose $\chi=0.06$ as a threshold for a predictive overlap. 

\begin{figure}[H]
	\centerline{\includegraphics[scale=0.6]{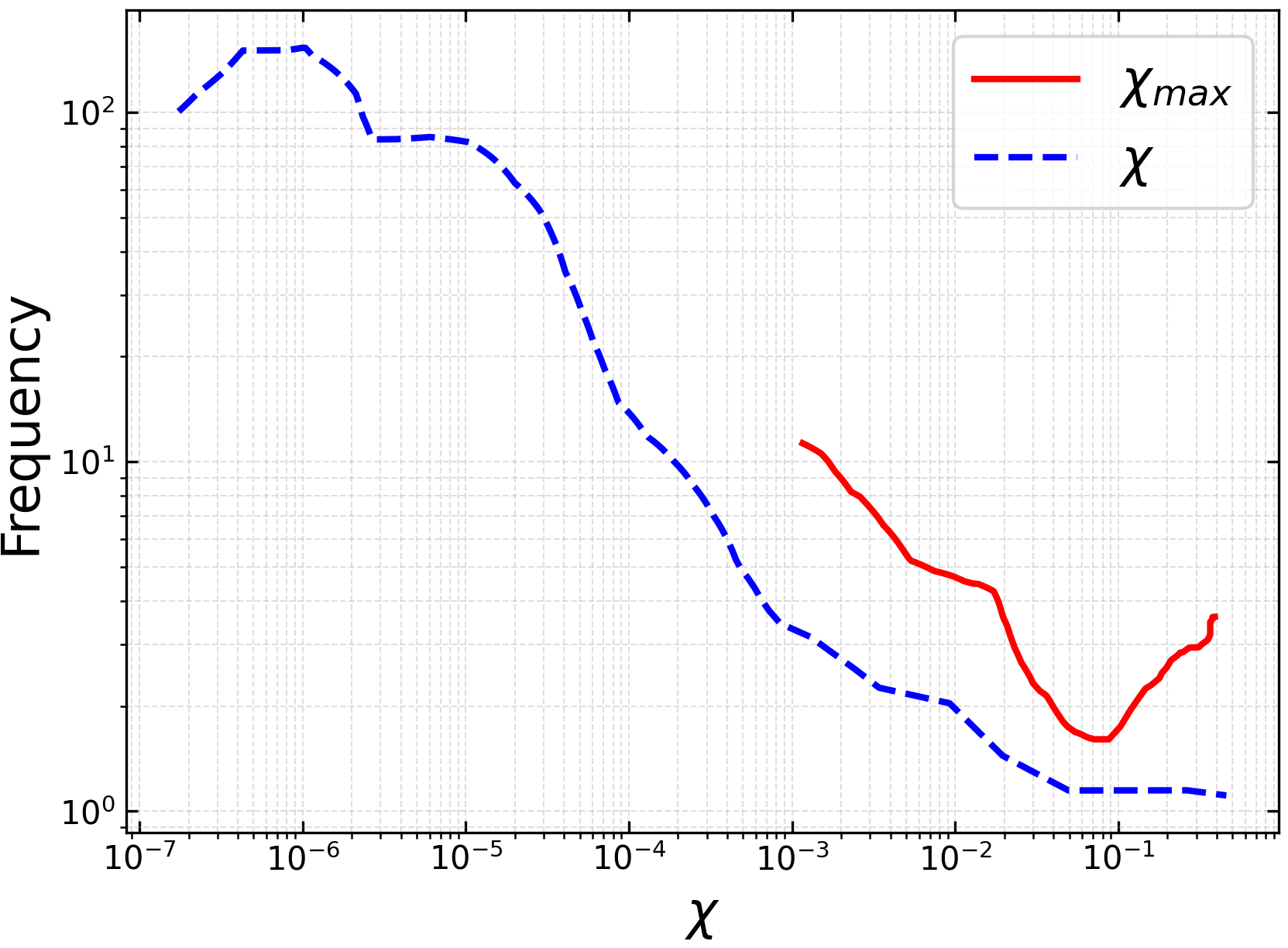}}
	\caption{\textbf{Decrease in predictive capacity when $\xiol =0.15$.} Distribution of proficiency values for all local shear clusters with all \dtwomin\, clusters. Blue dotted line shows $\chi$ for all pairs of clusters except $\chi_{max}$ for each \dtwomin\, cluster. Red line shows the distribution of $\chi_{max}$ for all \dtwomin\, clusters. }
	\label{fig:SI_prof_hist_xi=.15}
\end{figure}

Fig.~\ref{fig:SI_prof_hist_xi=.15} shows the same data for $\xiol = 0.15$, where the predictive rate is much lower (seen in Fig.~\ref{fig:6}). The distribution highlights this, as the majority of the maximum proficiencies are lower than the threshold (red curve), indicating that most of the overlaps are small and in the part of the bimodal distribution corresponding to the noise.

\bibliography{apssamp}

@PREAMBLE{
 "\providecommand{\noopsort}[1]{}" 
 # "\providecommand{\singleletter}[1]{#1}%" 
}

@article{ghaznavi2025yielding,
  title={Yielding in dense active matter [Dataset]},
  author={Ghaznavi, Adil and Rossi, Saverio and Zamponi, Francesco and Manning, M Lisa},
  journal={Dryad repository},
  DOI = {https://doi.org/10.5061/dryad.1ns1rn976},
  year={2026}
}

@article{rottler2018,
  title = {Orientation of plastic rearrangements in two-dimensional model glasses under shear},
  author = {Nicolas, Alexandre and Rottler, J\"org},
  journal = {Phys. Rev. E},
  volume = {97},
  issue = {6},
  pages = {063002},
  numpages = {9},
  year = {2018},
  month = {Jun},
  publisher = {American Physical Society},
  doi = {10.1103/PhysRevE.97.063002},
  url = {https://link.aps.org/doi/10.1103/PhysRevE.97.063002}
}

@article{giannini2022searching,
  title={Searching for structural predictors of plasticity in dense active packings},
  author={Giannini, Julia A and Stanifer, Ethan M and Manning, M Lisa},
  journal={Soft Matter},
  volume={18},
  number={7},
  pages={1540--1553},
  year={2022},
  publisher={Royal Society of Chemistry}
}

@article{hain2026,
  title={Using the force landscape of an active solid to predict plastic deformation},
  author={Hain, Tyler and Lerner, Edan and Manning, M Lisa},
  journal={arXiv preprint arXiv:2603.11425},
  year={2026}
}

@article{gandikota2026jammed,
  title={The Jammed Phase of Infinitely Persistent Active Matter},
  author={Gandikota, MC and Mandal, Rituparno and Chaudhuri, Pinaki and Chakraborty, Bulbul and Dasgupta, Chandan},
  journal={arXiv preprint arXiv:2602.20776},
  year={2026}
}

@article{wiese2025avalanches,
  title={Avalanches in active glasses with finite persistence},
  author={Wiese, Roland and Ferrero, Ezequiel and Levis, Demian},
  journal={arXiv preprint arXiv:2512.00523},
  year={2025}
}

@article{garcimartin2015flow,
  title={Flow and clogging of a sheep herd passing through a bottleneck},
  author={Garcimart{\'\i}n, A and Pastor, JM and Ferrer, LM and Ramos, JJ and Mart{\'\i}n-G{\'o}mez, C and Zuriguel, I},
  journal={Physical Review E},
  volume={91},
  number={2},
  pages={022808},
  year={2015},
  publisher={APS}
}

@article{ghosh2025elastoplastic,
  title={An elastoplastic model approach for the relaxation dynamics of active glasses},
  author={Ghosh, Tanmoy and Sollich, Peter and Nandi, Saroj Kumar},
  journal={Soft Matter},
  volume={21},
  number={16},
  pages={3047--3057},
  year={2025},
  publisher={Royal Society of Chemistry}
}

@article{berthier2013non,
  title={Non-equilibrium glass transitions in driven and active matter},
  author={Berthier, Ludovic and Kurchan, Jorge},
  journal={Nature Physics},
  volume={9},
  number={5},
  pages={310--314},
  year={2013},
  publisher={Nature Publishing Group UK London}
}

@article{berthier2017active,
  title={How active forces influence nonequilibrium glass transitions},
  author={Berthier, Ludovic and Flenner, Elijah and Szamel, Grzegorz},
  journal={New Journal of Physics},
  volume={19},
  number={12},
  pages={125006},
  year={2017},
  publisher={IOP Publishing}
}

@article{gu2025emergence,
  title={Emergence of collective oscillations in massive human crowds},
  author={Gu, Fran{\c{c}}ois and Guiselin, Benjamin and Bain, Nicolas and Zuriguel, Iker and Bartolo, Denis},
  journal={Nature},
  volume={638},
  number={8049},
  pages={112--119},
  year={2025},
  publisher={Nature Publishing Group UK London}
}

@article{Eshelby_1957, title={The determination of the elastic field of an ellipsoidal inclusion, and related problems}, volume={241}, DOI={10.1098/rspa.1957.0133}, abstractNote={It is supposed that a region within an isotropic elastic solid undergoes a spontaneous change of form which, if the surrounding material were absent, would be some prescribed homogeneous deformation. Because of the presence of the surrounding material stresses will be present both inside and outside the region. The resulting elastic field may be found very simply with the help of a sequence of imaginary cutting, straining and welding operations. In particular, if the region is an ellipsoid the strain inside it is uniform and may be expressed in terms of tabu­lated elliptic integrals. In this case a further problem may be solved. An ellipsoidal region in an infinite medium has elastic constants different from those of the rest of the material; how does the presence of this inhomogeneity disturb an applied stress-field uniform at large distances? It is shown that to answer several questions of physical or engineering interest it is necessary to know only the relatively simple elastic field inside the ellipsoid.}, number={1226}, journal={Proceedings of the Royal Society of London. Series A, Mathematical and physical sciences}, publisher={Royal Society, The}, author={Eshelby, John Douglas}, year={1957}, month=aug, pages={376–396} }

@article{
pareek2026,
author = {Puneet Pareek  and Peter Sollich  and Saroj Kumar Nandi  and Ludovic Berthier },
title = {Scaling the glassy dynamics of active particles: Tunable fragility and reentrance},
journal = {Proceedings of the National Academy of Sciences},
volume = {123},
number = {4},
pages = {e2516624123},
year = {2026},
doi = {10.1073/pnas.2516624123},
URL = {https://www.pnas.org/doi/abs/10.1073/pnas.2516624123},
eprint = {https://www.pnas.org/doi/pdf/10.1073/pnas.2516624123}}

@article{pierce_hydrodynamic_2018,
	title = {Hydrodynamic {Interactions}, {Hidden} {Order}, and {Emergent} {Collective} {Behavior} in an {Active} {Bacterial} {Suspension}},
	volume = {121},
	issn = {0031-9007, 1079-7114},
	url = {https://link.aps.org/doi/10.1103/PhysRevLett.121.188001},
	doi = {10.1103/PhysRevLett.121.188001},
	number = {18},
	urldate = {2021-10-25},
	journal = {Physical Review Letters},
	author = {Pierce, C. J. and Wijesinghe, H. and Mumper, E. and Lower, B. H. and Lower, S. K. and Sooryakumar, R.},
	month = nov,
	year = {2018},
	pages = {188001},
}

@article{schotz_glassy_2013,
	title = {Glassy dynamics in three-dimensional embryonic tissues},
	volume = {10},
	issn = {1742-5689},
	url = {https://www.ncbi.nlm.nih.gov/pmc/articles/PMC3808551/},
	doi = {10.1098/rsif.2013.0726},
	number = {89},
	urldate = {2020-07-16},
	journal = {Journal of the Royal Society Interface},
	author = {Schötz, Eva-Maria and Lanio, Marcos and Talbot, Jared A. and Manning, M. Lisa},
	month = dec,
	pages = {1--11},
	year = {2013},
	pmid = {24068179},
	pmcid = {PMC3808551},
}

@article{cavagna_empirical_2010,
	title = {From empirical data to inter-individual interactions: unveiling the rules of collective animal behavior},
	volume = {20},
	issn = {0218-2025, 1793-6314},
	shorttitle = {{FROM} {EMPIRICAL} {DATA} {TO} {INTER}-{INDIVIDUAL} {INTERACTIONS}},
	url = {https://www.worldscientific.com/doi/abs/10.1142/S0218202510004660},
	doi = {10.1142/S0218202510004660},
	number = {supp01},
	urldate = {2021-10-25},
	journal = {Mathematical Models and Methods in Applied Sciences},
	author = {Cavagna, Andrea and Cimarelli, Alessio and Giardina, Irene and Parisi, Giorgio and Santagati, Raffaele and Stefanini, Fabio and Tavarone, Raffaele},
	month = sep,
	year = {2010},
	pages = {1491--1510},
}

@article{marchetti_hydrodynamics_2013,
	title = {Hydrodynamics of soft active matter},
	volume = {85},
	issn = {0034-6861, 1539-0756},
	url = {https://link.aps.org/doi/10.1103/RevModPhys.85.1143},
	doi = {10.1103/RevModPhys.85.1143},
	number = {3},
	urldate = {2021-10-25},
	journal = {Reviews of Modern Physics},
	author = {Marchetti, M. C. and Joanny, J. F. and Ramaswamy, S. and Liverpool, T. B. and Prost, J. and Rao, Madan and Simha, R. Aditi},
	month = jul,
	year = {2013},
	pages = {1143--1189},
}

@article{Salerno2012,
  doi = {10.1103/physrevlett.109.105703},
  url = {https://doi.org/10.1103/physrevlett.109.105703},
  year = {2012},
  month = sep,
  publisher = {American Physical Society ({APS})},
  volume = {109},
  number = {10},
  pages = {105703},
  author = {K. Michael Salerno and Craig E. Maloney and Mark O. Robbins},
  title = {Avalanches in Strained Amorphous Solids: Does Inertia Destroy Critical Behavior?},
  journal = {Physical Review Letters}
}

@article{morse_direct_2021,
	title = {A direct link between active matter and sheared granular systems},
	volume = {118},
	url = {https://www.pnas.org/doi/full/10.1073/pnas.2019909118},
	doi = {10.1073/pnas.2019909118},
	pages = {e2019909118},
	number = {18},
	journaltitle = {Proceedings of the National Academy of Sciences},
	author = {Morse, Peter K. and Roy, Sudeshna and Agoritsas, Elisabeth and Stanifer, Ethan and Corwin, Eric I. and Manning, M. Lisa},
	urldate = {2022-05-11},
	date = {2021-05-04},
	note = {Publisher: Proceedings of the National Academy of Sciences},
}

@article{brito_theory_2018,
	title = {Theory for Swap Acceleration near the Glass and Jamming Transitions for Continuously Polydisperse Particles},
	volume = {8},
	issn = {2160-3308},
	url = {https://link.aps.org/doi/10.1103/PhysRevX.8.031050},
	doi = {10.1103/PhysRevX.8.031050},
	pages = {031050},
	number = {3},
	journaltitle = {Physical Review X},
	shortjournal = {Phys. Rev. X},
	author = {Brito, Carolina and Lerner, Edan and Wyart, Matthieu},
	urldate = {2022-10-22},
	date = {2018-08-27},
	langid = {english},
}

@article{richard_predicting_2020,
	title = {Predicting plasticity in disordered solids from structural indicators},
	volume = {4},
	issn = {2475-9953},
	url = {http://arxiv.org/abs/2003.11629},
	doi = {10.1103/PhysRevMaterials.4.113609},
	pages = {113609},
	number = {11},
	journaltitle = {Physical Review Materials},
	shortjournal = {Phys. Rev. Materials},
	author = {Richard, D. and Ozawa, M. and Patinet, S. and Stanifer, E. and Shang, B. and Ridout, S. A. and Xu, B. and Zhang, G. and Morse, P. K. and Barrat, J.-L. and Berthier, L. and Falk, M. L. and Guan, P. and Liu, A. J. and Martens, K. and Sastry, S. and Vandembroucq, D. and Lerner, E. and Manning, M. L.},
	urldate = {2022-09-08},
	date = {2020-11-24},
	eprinttype = {arxiv},
	eprint = {2003.11629 [cond-mat]},
}

@article{nicolas_deformation_2018,
	title = {Deformation and flow of amorphous solids: An updated review of mesoscale elastoplastic models},
	volume = {90},
	issn = {0034-6861, 1539-0756},
	url = {http://arxiv.org/abs/1708.09194},
	doi = {10.1103/RevModPhys.90.045006},
	shorttitle = {Deformation and flow of amorphous solids},
	pages = {045006},
	number = {4},
	journaltitle = {Reviews of Modern Physics},
	shortjournal = {Rev. Mod. Phys.},
	author = {Nicolas, Alexandre and Ferrero, Ezequiel E. and Martens, Kirsten and Barrat, Jean-Louis},
	urldate = {2023-04-02},
	date = {2018-12-26},
	eprinttype = {arxiv},
	eprint = {1708.09194 [cond-mat]},
}

@article{pollard_yielding_2022,
	title = {Yielding, shear banding and brittle failure of amorphous materials},
	volume = {4},
	issn = {2643-1564},
	url = {http://arxiv.org/abs/2103.06782},
	doi = {10.1103/PhysRevResearch.4.043037},
	pages = {043037},
	number = {4},
	journaltitle = {Physical Review Research},
	shortjournal = {Phys. Rev. Research},
	author = {Pollard, Joseph and Fielding, Suzanne M.},
	urldate = {2023-04-25},
	date = {2022-10-17},
	eprinttype = {arxiv},
	eprint = {2103.06782 [cond-mat]},
}

@article{agoritsas_mean-field_2021,
	title = {Mean-field dynamics of infinite-dimensional particle systems: global shear versus random local forcing},
	volume = {2021},
	issn = {1742-5468},
	url = {http://arxiv.org/abs/2009.08944},
	doi = {10.1088/1742-5468/abdd18},
	shorttitle = {Mean-field dynamics of infinite-dimensional particle systems},
	pages = {033501},
	number = {3},
	journaltitle = {Journal of Statistical Mechanics: Theory and Experiment},
	shortjournal = {J. Stat. Mech.},
	author = {Agoritsas, Elisabeth},
	urldate = {2023-07-06},
	date = {2021-03-01},
	eprinttype = {arxiv},
	eprint = {2009.08944 [cond-mat]},
}

@article{Hagh_2022,
	doi = {10.1073/pnas.2117622119},
  
	url = {https://doi.org/10.1073%2Fpnas.2117622119},
  
	year = 2022,
	month = {may},
  
	publisher = {Proceedings of the National Academy of Sciences},
  
	volume = {119},
  
	number = {19},
  
	author = {Varda F. Hagh and Sidney R. Nagel and Andrea J. Liu and M. Lisa Manning and Eric I. Corwin},
  
	title = {Transient learning degrees of freedom for introducing function in materials},
  
	journal = {Proceedings of the National Academy of Sciences}
}

@article{rainone_following_2016,
	title = {Following the evolution of glassy states under external perturbations: the full replica symmetry breaking solution},
	volume = {2016},
	issn = {1742-5468},
	shorttitle = {Following the evolution of glassy states under external perturbations},
	url = {https://doi.org/10.1088%2F1742-5468%2F2016%2F05%2F053302},
	doi = {10.1088/1742-5468/2016/05/053302},
	language = {en},
	number = {5},
	urldate = {2019-12-05},
	journal = {J. Stat. Mech.},
	author = {Rainone, Corrado and Urbani, Pierfrancesco},
	month = may,
	year = {2016},
	pages = {053302},
}

@article{biroli_breakdown_2016,
	title = {Breakdown of {Elasticity} in {Amorphous} {Solids}},
	volume = {12},
	issn = {1745-2473, 1745-2481},
	url = {http://arxiv.org/abs/1601.06724},
	doi = {10.1038/nphys3845},
	number = {12},
	urldate = {2020-09-21},
	journal = {Nature Phys},
	author = {Biroli, Giulio and Urbani, Pierfrancesco},
	month = dec,
	year = {2016},
	pages = {1130--1133},
}

@article{urbani_shear_2017,
	title = {Shear {Yielding} and {Shear} {Jamming} of {Dense} {Hard} {Sphere} {Glasses}},
	volume = {118},
	url = {https://link.aps.org/doi/10.1103/PhysRevLett.118.038001},
	doi = {10.1103/PhysRevLett.118.038001},
	number = {3},
	urldate = {2018-05-08},
	journal = {Phys. Rev. Lett.},
	author = {Urbani, Pierfrancesco and Zamponi, Francesco},
	month = jan,
	year = {2017},
	pages = {038001},
}

@article{biroli_liu-nagel_2018,
	title = {Liu-{Nagel} phase diagrams in infinite dimension},
	volume = {4},
	issn = {2542-4653},
	url = {http://arxiv.org/abs/1704.04649},
	doi = {10.21468/SciPostPhys.4.4.020},
	number = {4},
	urldate = {2020-09-21},
	journal = {SciPost Phys.},
	author = {Biroli, Giulio and Urbani, Pierfrancesco},
	month = apr,
	year = {2018},
	pages = {020},
}

@article{stanifer_avalanche_2022,
	title = {Avalanche dynamics in sheared athermal particle packings occurs via localized bursts predicted by unstable linear response},
	url = {http://arxiv.org/abs/2110.02803},
	language = {en},
	urldate = {2024-06-18},
	publisher = {arXiv},
	author = {Stanifer, Ethan and Manning, M. Lisa},
	month = mar,
	year = {2022},
	note = {arXiv:2110.02803 [cond-mat]},
}

@article{altieri_mean-field_2019,
	title = {Mean-field stability map of hard-sphere glasses},
	volume = {100},
	url = {https://link.aps.org/doi/10.1103/PhysRevE.100.032140},
	doi = {10.1103/PhysRevE.100.032140},
	number = {3},
	urldate = {2020-08-11},
	journal = {Phys. Rev. E},
	author = {Altieri, Ada and Zamponi, Francesco},
	month = sep,
	year = {2019},
	pages = {032140},
}

@article{Rossi2022,
  title={Finite-disorder critical point in the yielding transition of elastoplastic models},
  author={Rossi, Saverio and Biroli, Giulio and Ozawa, Misaki and Tarjus, Gilles and Zamponi, Francesco},
  journal={Phys. Rev. Lett.},
  volume={129},
  number={22},
  pages={228002},
  year={2022},
  publisher={APS}
}

@article{manning2011vibrational,
  title={Vibrational modes identify soft spots in a sheared disordered packing},
  author={Manning, M Lisa and Liu, Andrea J},
  journal={Physical Review Letters},
  volume={107},
  number={10},
  pages={108302},
  year={2011},
  publisher={APS}
}

@article{falk1998dynamics,
  title={Dynamics of viscoplastic deformation in amorphous solids},
  author={Falk, ML and Langer, JS},
  journal={Physical Review E},
  volume={57},
  number={6},
  pages={7192},
  year={1998},
  publisher={APS}
}

@article{schall2007structural,
  title={Structural rearrangements that govern flow in colloidal glasses},
  author={Schall, Peter and Weitz, David A and Spaepen, Frans},
  journal={Science},
  volume={318},
  number={5858},
  pages={1895--1899},
  year={2007},
  publisher={American Association for the Advancement of Science}
}

@article{bottinelli2016emergent,
  title={Emergent structural mechanisms for high-density collective motion inspired by human crowds},
  author={Bottinelli, Arianna and Sumpter, David TJ and Silverberg, Jesse L},
  journal={Physical review letters},
  volume={117},
  number={22},
  pages={228301},
  year={2016},
  publisher={APS}
}

@incollection{sollich2006soft,
  title={Soft glassy rheology},
  author={Sollich, Peter},
  booktitle={Molecular Gels},
  pages={161--192},
  year={2006},
  publisher={Springer}
}

@article{OHern2003,
  doi = {10.1103/physreve.68.011306},
  url = {https://doi.org/10.1103/physreve.68.011306},
  year = {2003},
  month = jul,
  publisher = {American Physical Society ({APS})},
  volume = {68},
  number = {1},
  pages={011306},
  author = {Corey S. O'Hern and Leonardo E. Silbert and Andrea J. Liu and Sidney R. Nagel},
  title = {Jamming at zero temperature and zero applied stress: The epitome of disorder},
  journal = {Physical Review E}
}

@article{henkes_active_2011,
	title = {Active {Jamming}: {Self}-propelled soft particles at high density},
	volume = {84},
	issn = {1539-3755, 1550-2376},
	shorttitle = {Active {Jamming}},
	url = {http://arxiv.org/abs/1107.4072},
	doi = {10.1103/PhysRevE.84.040301},
	number = {4},
	urldate = {2018-05-01},
	journal = {Physical Review E},
	author = {Henkes, Silke and Fily, Yaouen and Marchetti, M. Cristina},
	month = oct,
	year = {2011},
	note = {arXiv: 1107.4072}
}

@article{bi_motility-driven_2016,
	title = {Motility-{Driven} {Glass} and {Jamming} {Transitions} in {Biological} {Tissues}},
	volume = {6},
	url = {https://link.aps.org/doi/10.1103/PhysRevX.6.021011},
	doi = {10.1103/PhysRevX.6.021011},
	number = {2},
	urldate = {2018-05-01},
	journal = {Physical Review X},
	author = {Bi, Dapeng and Yang, Xingbo and Marchetti, M. Cristina and Manning, M. Lisa},
	month = apr,
	year = {2016},
	pages = {021011}
}

@article{vicsek_novel_1995,
	title = {Novel {Type} of {Phase} {Transition} in a {System} of {Self}-{Driven} {Particles}},
	volume = {75},
	url = {https://link.aps.org/doi/10.1103/PhysRevLett.75.1226},
	doi = {10.1103/PhysRevLett.75.1226},
	number = {6},
	urldate = {2020-05-01},
	journal = {Physical Review Letters},
	author = {Vicsek, Tamás and Czirók, András and Ben-Jacob, Eshel and Cohen, Inon and Shochet, Ofer},
	month = aug,
	year = {1995},
	note = {Publisher: American Physical Society},
	pages = {1226--1229}
}

@article{malandro1998molecular,
  title={Molecular-level mechanical instabilities and enhanced self-diffusion in flowing liquids},
  author={Malandro, Dennis L and Lacks, Daniel J},
  journal={Physical review letters},
  volume={81},
  number={25},
  pages={5576},
  year={1998},
  publisher={APS}
}

@article{utz2000atomistic,
  title={Atomistic simulation of aging and rejuvenation in glasses},
  author={Utz, Marcel and Debenedetti, Pablo G and Stillinger, Frank H},
  journal={Physical review letters},
  volume={84},
  number={7},
  pages={1471},
  year={2000},
  publisher={APS}
}

@article{singh2020brittle,
  title={Brittle yielding of amorphous solids at finite shear rates},
  author={Singh, Murari and Ozawa, Misaki and Berthier, Ludovic},
  journal={Physical Review Materials},
  volume={4},
  number={2},
  pages={025603},
  year={2020},
  publisher={APS}
}

@article{schuh2003atomistic,
  title={Atomistic basis for the plastic yield criterion of metallic glass},
  author={Schuh, Christopher A and Lund, Alan C},
  journal={Nature materials},
  volume={2},
  number={7},
  pages={449--452},
  year={2003},
  publisher={Nature Publishing Group UK London}
}

@article{utz2004athermal,
  title={Athermal simulation of plastic deformation in amorphous solids at constant pressure},
  author={Utz, Marcel and Peng, Qing and Nandagopal, Magesh},
  journal={Journal of Polymer Science Part B: Polymer Physics},
  volume={42},
  number={11},
  pages={2057--2065},
  year={2004},
  publisher={Wiley Online Library}
}

@article{maloney_amorphous_2006,
	title = {Amorphous systems in athermal, quasistatic shear},
	volume = {74},
	url = {https://link.aps.org/doi/10.1103/PhysRevE.74.016118},
	doi = {10.1103/PhysRevE.74.016118},
	number = {1},
	journal = {Physical Review E},
	author = {Maloney, Craig E. and Lemaître, Anaël},
	month = jul,
	year = {2006},
	pages = {016118}
}

@article{Patinet2016,
  doi = {10.1103/physrevlett.117.045501},
  url = {https://doi.org/10.1103/physrevlett.117.045501},
  year = {2016},
  month = jul,
  publisher = {American Physical Society ({APS})},
  volume = {117},
  number = {4},
  pages ={045501},
  author = {Sylvain Patinet and Damien Vandembroucq and Michael L. Falk},
  title = {Connecting Local Yield Stresses with Plastic Activity in Amorphous Solids},
  journal = {Physical Review Letters}
}

@article{Popovi2018,
  doi = {10.1103/physreve.98.040901},
  url = {https://doi.org/10.1103/physreve.98.040901},
  year = {2018},
  month = oct,
  publisher = {American Physical Society ({APS})},
  volume = {98},
  number = {4},
  pages = {040901},
  author = {Marko Popovi{\'{c}} and Tom W. J. de Geus and Matthieu Wyart},
  title = {Elastoplastic description of sudden failure in athermal amorphous materials during quasistatic loading},
  journal = {Physical Review E}
}

@article{Ozawa2018,
  doi = {10.1073/pnas.1806156115},
  url = {https://doi.org/10.1073/pnas.1806156115},
  year = {2018},
  month = jun,
  publisher = {Proceedings of the National Academy of Sciences},
  volume = {115},
  number = {26},
  pages = {6656--6661},
  author = {Misaki Ozawa and Ludovic Berthier and Giulio Biroli and Alberto Rosso and Gilles Tarjus},
  title = {Random critical point separates brittle and ductile yielding transitions in amorphous materials},
  journal = {Proceedings of the National Academy of Sciences}
}

@article{gartner2016nonlinear,
  title={Nonlinear modes disentangle glassy and Goldstone modes in structural glasses},
  author={Gartner, Luka and Lerner, Edan and others},
  journal={SciPost Phys},
  volume={1},
  number={2},
  pages={016},
  year={2016}
}

@book{johnson1987contact,
  title={Contact mechanics},
  author={Johnson, Kenneth Langstreth and Johnson, Kenneth Langstreth},
  year={1987},
  publisher={Cambridge university press}
}

@article{nicolas2014,
  title={Spatiotemporal correlations between plastic events in the shear flow of athermal amorphous solids},
  author={Nicolas, Alexandre and Rottler, Joerg and Barrat, Jean-Louis},
  journal={The European Physical Journal E},
  volume={37},
  number={6},
  pages={1--11},
  year={2014},
  publisher={Springer}
}

@article{lagogianni2018,
  title={Plastic avalanches in the so-called elastic regime of metallic glasses},
  author={Lagogianni, Alexandra E and Liu, Chen and Martens, Kirsten and Samwer, Konrad},
  journal={The European Physical Journal B},
  volume={91},
  number={6},
  pages={1--5},
  year={2018},
  publisher={Springer}
}

@article{zhang2020interplay,
  title={Interplay of rearrangements, strain, and local structure during avalanche propagation},
  author={Zhang, Ge and Ridout, Sean and Liu, Andrea J},
  journal={arXiv preprint arXiv:2009.11414},
  year={2020}
}

@article{yeomans2025active,
  title={Active nematics: a new approach to mechanobiology?},
  author={Yeomans, Julia M and Bhattacharyya, Saraswat and Nejad, Mehrana R},
  journal={Liquid Crystals},
  pages={1--9},
  year={2025},
  publisher={Taylor \& Francis}
}

@article{kempf2019active,
  title={Active matter invasion},
  author={Kempf, Felix and Mueller, Romain and Frey, Erwin and Yeomans, Julia M and Doostmohammadi, Amin},
  journal={Soft matter},
  volume={15},
  number={38},
  pages={7538--7546},
  year={2019},
  publisher={Royal Society of Chemistry}
}

@article{foster2022active,
  title={Active mechanics of sea star oocytes},
  author={Foster, Peter J and F{\"u}rthauer, Sebastian and Fakhri, Nikta},
  journal={BioRxiv},
  pages={2022--04},
  year={2022},
  publisher={Cold Spring Harbor Laboratory}
}

@article{toner2018swarming,
  title={Swarming in the dirt: Ordered flocks with quenched disorder},
  author={Toner, John and Guttenberg, Nicholas and Tu, Yuhai},
  journal={Physical review letters},
  volume={121},
  number={24},
  pages={248002},
  year={2018},
  publisher={APS}
}

@article{fruchart2023odd,
  title={Odd viscosity and odd elasticity},
  author={Fruchart, Michel and Scheibner, Colin and Vitelli, Vincenzo},
  journal={Annual Review of Condensed Matter Physics},
  volume={14},
  number={1},
  pages={471--510},
  year={2023},
  publisher={Annual Reviews}
}

@article{nonrecippopa2014,
  title={Non-reciprocal and highly nonlinear active acoustic metamaterials},
  author={Popa, Bogdan-Ioan and Cummer, Steven A},
  journal={Nature communications},
  volume={5},
  number={1},
  pages={3398},
  year={2014},
  publisher={Nature Publishing Group UK London}
}

@article{ishimoto2023odd,
  title={Odd elastohydrodynamics: non-reciprocal living material in a viscous fluid},
  author={Ishimoto, Kenta and Moreau, Cl{\'e}ment and Yasuda, Kento},
  journal={PRX Life},
  volume={1},
  number={2},
  pages={023002},
  year={2023},
  publisher={APS}
}

@article{tonertu1995XY,
  title = {Long-Range Order in a Two-Dimensional Dynamical $\mathrm{XY}$ Model: How Birds Fly Together},
  author = {Toner, John and Tu, Yuhai},
  journal = {Phys. Rev. Lett.},
  volume = {75},
  issue = {23},
  pages = {4326--4329},
  numpages = {0},
  year = {1995},
  month = {Dec},
  publisher = {American Physical Society},
  doi = {10.1103/PhysRevLett.75.4326},
  url = {https://link.aps.org/doi/10.1103/PhysRevLett.75.4326}
}

@article{tonertu1998flocking,
  title = {Flocks, herds, and schools: A quantitative theory of flocking},
  author = {Toner, John and Tu, Yuhai},
  journal = {Phys. Rev. E},
  volume = {58},
  issue = {4},
  pages = {4828--4858},
  numpages = {0},
  year = {1998},
  month = {Oct},
  publisher = {American Physical Society},
  doi = {10.1103/PhysRevE.58.4828},
  url = {https://link.aps.org/doi/10.1103/PhysRevE.58.4828}
}

@article{TONERturamaswamy2005170flocks,
title = {Hydrodynamics and phases of flocks},
journal = {Annals of Physics},
volume = {318},
number = {1},
pages = {170-244},
year = {2005},
note = {Special Issue},
issn = {0003-4916},
doi = {https://doi.org/10.1016/j.aop.2005.04.011},
url = {https://www.sciencedirect.com/science/article/pii/S0003491605000540},
author = {John Toner and Yuhai Tu and Sriram Ramaswamy}
}

@article{cavagna2014bird,
  title={Bird flocks as condensed matter},
  author={Cavagna, Andrea and Giardina, Irene},
  journal={Annu. Rev. Condens. Matter Phys.},
  volume={5},
  number={1},
  pages={183--207},
  year={2014},
  publisher={Annual Reviews}
}

@article{cavagna2015flocking,
  title={Flocking and turning: a new model for self-organized collective motion},
  author={Cavagna, Andrea and Del Castello, Lorenzo and Giardina, Irene and Grigera, Tomas and Jelic, Asja and Melillo, Stefania and Mora, Thierry and Parisi, Leonardo and Silvestri, Edmondo and Viale, Massimiliano and others},
  journal={Journal of Statistical Physics},
  volume={158},
  number={3},
  pages={601--627},
  year={2015},
  publisher={Springer}
}

@article{mandal2020extreme,
  title={Extreme active matter at high densities},
  author={Mandal, Rituparno and Bhuyan, Pranab Jyoti and Chaudhuri, Pinaki and Dasgupta, Chandan and Rao, Madan},
  journal={Nature communications},
  volume={11},
  number={1},
  pages={2581},
  year={2020},
  publisher={Nature Publishing Group UK London}
}

@article{mandal2021rheology,
  title={Rheology of cohesive granular media: Shear banding, hysteresis, and nonlocal effects},
  author={Mandal, Sandip and Nicolas, Maxime and Pouliquen, Olivier},
  journal={Physical Review X},
  volume={11},
  number={2},
  pages={021017},
  year={2021},
  publisher={APS}
}

@article{woillez2020active,
  title={Active trap model},
  author={Woillez, Eric and Kafri, Yariv and Gov, Nir S},
  journal={Physical Review Letters},
  volume={124},
  number={11},
  pages={118002},
  year={2020},
  publisher={APS}
}

@article{liao2018criticality,
  title={Criticality of the zero-temperature jamming transition probed by self-propelled particles},
  author={Liao, Qinyi and Xu, Ning},
  journal={Soft Matter},
  volume={14},
  number={5},
  pages={853--860},
  year={2018},
  publisher={Royal Society of Chemistry}
}

@article{berthier2025yielding,
  title={Yielding and plasticity in amorphous solids},
  author={Berthier, Ludovic and Biroli, Giulio and Manning, Lisa and Zamponi, Francesco},
  journal={Nature Reviews Physics},
  pages={1--18},
  year={2025},
  publisher={Nature Publishing Group UK London}
}

@article{olmsted2008perspectives,
  title={Perspectives on shear banding in complex fluids},
  author={Olmsted, Peter D},
  journal={Rheologica Acta},
  volume={47},
  number={3},
  pages={283--300},
  year={2008},
  publisher={Springer}
}

@article{fieldingcates2011nonlinear,
  title={Nonlinear dynamics and rheology of active fluids: Simulations in two dimensions},
  author={Fielding, Suzanne M and Marenduzzo, Davide and Cates, Michael E},
  journal={Physical Review E—Statistical, Nonlinear, and Soft Matter Physics},
  volume={83},
  number={4},
  pages={041910},
  year={2011},
  publisher={APS}
}

@article{catesfielding2006rheology,
  title={Rheology of giant micelles},
  author={Cates, Michael E and Fielding, Suzanne M},
  journal={Advances in Physics},
  volume={55},
  number={7-8},
  pages={799--879},
  year={2006},
  publisher={Taylor \& Francis}
}

@article{fielding2007complex,
  title={Complex dynamics of shear banded flows},
  author={Fielding, Suzanne M},
  journal={Soft Matter},
  volume={3},
  number={10},
  pages={1262--1279},
  year={2007},
  publisher={Royal Society of Chemistry}
}

@article{wyartlernerdegiuli2015theory,
  title={Theory of the jamming transition at finite temperature},
  author={Degiuli, Eric and Lerner, E and Wyart, M},
  journal={The Journal of chemical physics},
  volume={142},
  number={16},
  year={2015},
  publisher={AIP Publishing}
}

@article{coussot2002avalanche,
  title={Avalanche behavior in yield stress fluids},
  author={Coussot, Philippe and Nguyen, Quoc Dzuy and Huynh, HT and Bonn, Daniel},
  journal={Physical review letters},
  volume={88},
  number={17},
  pages={175501},
  year={2002},
  publisher={APS}
}

@article{baro2018universalavalanche,
  title={Universal avalanche statistics and triggering close to failure in a mean-field model of rheological fracture},
  author={Bar{\'o}, Jordi and Davidsen, J{\"o}rn},
  journal={Physical Review E},
  volume={97},
  number={3},
  pages={033002},
  year={2018},
  publisher={APS}
}

@article{dahmen2011simpleavalanche,
  title={A simple analytic theory for the statistics of avalanches in sheared granular materials},
  author={Dahmen, Karin A and Ben-Zion, Yehuda and Uhl, Jonathan T},
  journal={Nature Physics},
  volume={7},
  number={7},
  pages={554--557},
  year={2011},
  publisher={Nature Publishing Group UK London}
}

@article{castellanos2018avalanche,
  title={Avalanche behavior in creep failure of disordered materials},
  author={Castellanos, David Fernandez and Zaiser, Michael},
  journal={Physical review letters},
  volume={121},
  number={12},
  pages={125501},
  year={2018},
  publisher={APS}
}

@article{karimi2017inertiaavalanche,
  title={Inertia and universality of avalanche statistics: The case of slowly deformed amorphous solids},
  author={Karimi, Kamran and Ferrero, Ezequiel E and Barrat, Jean-Louis},
  journal={Physical Review E},
  volume={95},
  number={1},
  pages={013003},
  year={2017},
  publisher={APS}
}

@article{fieldingdivoux2024ductile,
  title={Ductile-to-brittle transition and yielding in soft amorphous materials: perspectives and open questions},
  author={Divoux, Thibaut and Agoritsas, Elisabeth and Aime, Stefano and Barentin, Catherine and Barrat, Jean-Louis and Benzi, Roberto and Berthier, Ludovic and Bi, Dapeng and Biroli, Giulio and Bonn, Daniel and others},
  journal={Soft Matter},
  volume={20},
  number={35},
  pages={6868--6888},
  year={2024},
  publisher={Royal Society of Chemistry}
}

@article{fieldingbarlow2020ductile,
  title={Ductile and brittle yielding in thermal and athermal amorphous materials},
  author={Barlow, Hugh J and Cochran, James O and Fielding, Suzanne M},
  journal={Physical Review Letters},
  volume={125},
  number={16},
  pages={168003},
  year={2020},
  publisher={APS}
}

@article{kapteijns2019fast_breathing,
  title={Fast generation of ultrastable computer glasses by minimization of an augmented potential energy},
  author={Kapteijns, Geert and Ji, Wencheng and Brito, Carolina and Wyart, Matthieu and Lerner, Edan},
  journal={Physical Review E},
  volume={99},
  number={1},
  pages={012106},
  year={2019},
  publisher={APS}
}

@article{ninarello2017models_breathing,
  title={Models and algorithms for the next generation of glass transition studies},
  author={Ninarello, Andrea and Berthier, Ludovic and Coslovich, Daniele},
  journal={Physical Review X},
  volume={7},
  number={2},
  pages={021039},
  year={2017},
  publisher={APS}
}

@article{otter2017roadmap_persistenthomology,
  title={A roadmap for the computation of persistent homology},
  author={Otter, Nina and Porter, Mason A and Tillmann, Ulrike and Grindrod, Peter and Harrington, Heather A},
  journal={EPJ Data Science},
  volume={6},
  number={1},
  pages={17},
  year={2017},
  publisher={Springer}
}

@article{dell2018growing_bacterial,
  title={A growing bacterial colony in two dimensions as an active nematic},
  author={Dell’Arciprete, Dario and Blow, Matthew L and Brown, Aidan T and Farrell, Fred DC and Lintuvuori, Juho S and McVey, Alexander F and Marenduzzo, Davide and Poon, Wilson CK},
  journal={Nature communications},
  volume={9},
  number={1},
  pages={4190},
  year={2018},
  publisher={Nature Publishing Group UK London}
}

@article{bililign2022motile,
  title={Motile dislocations knead odd crystals into whorls},
  author={Bililign, Ephraim S and Balboa Usabiaga, Florencio and Ganan, Yehuda A and Poncet, Alexis and Soni, Vishal and Magkiriadou, Sofia and Shelley, Michael J and Bartolo, Denis and Irvine, William TM},
  journal={Nature Physics},
  volume={18},
  number={2},
  pages={212--218},
  year={2022},
  publisher={Nature Publishing Group UK London}
}

@article{cohen2014galvanotactic,
  title={Galvanotactic control of collective cell migration in epithelial monolayers},
  author={Cohen, Daniel J and James Nelson, W and Maharbiz, Michel M},
  journal={Nature materials},
  volume={13},
  number={4},
  pages={409--417},
  year={2014},
  publisher={Nature Publishing Group UK London}
}

@article{cohen2014emergent,
  title={Emergent cometlike swarming of optically driven thermally active colloids},
  author={Cohen, Jack A and Golestanian, Ramin},
  journal={Physical review letters},
  volume={112},
  number={6},
  pages={068302},
  year={2014},
  publisher={APS}
}

@article{wyart2014discontinuous,
  title={Discontinuous shear thickening without inertia in dense non-Brownian suspensions},
  author={Wyart, Matthieu and Cates, Micheal E},
  journal={Physical review letters},
  volume={112},
  number={9},
  pages={098302},
  year={2014},
  publisher={APS}
}

@article{janssen2019active,
  title={Active glasses},
  author={Janssen, Liesbeth MC},
  journal={Journal of Physics: Condensed Matter},
  volume={31},
  number={50},
  pages={503002},
  year={2019},
  publisher={IOP Publishing}
}

@article{paul2023dynamical,
  title={Dynamical heterogeneity in active glasses is inherently different from its equilibrium behavior},
  author={Paul, Kallol and Mutneja, Anoop and Nandi, Saroj Kumar and Karmakar, Smarajit},
  journal={Proceedings of the National Academy of Sciences},
  volume={120},
  number={34},
  pages={e2217073120},
  year={2023},
  publisher={National Academy of Sciences}
}

@article{nandi2018random,
  title={A random first-order transition theory for an active glass},
  author={Nandi, Saroj Kumar and Mandal, Rituparno and Bhuyan, Pranab Jyoti and Dasgupta, Chandan and Rao, Madan and Gov, Nir S},
  journal={Proceedings of the National Academy of Sciences},
  volume={115},
  number={30},
  pages={7688--7693},
  year={2018},
  publisher={National Academy of Sciences}
}

@article{henkes2020dense,
  title={Dense active matter model of motion patterns in confluent cell monolayers},
  author={Henkes, Silke and Kostanjevec, Kaja and Collinson, J Martin and Sknepnek, Rastko and Bertin, Eric},
  journal={Nature communications},
  volume={11},
  number={1},
  pages={1405},
  year={2020},
  publisher={Nature Publishing Group UK London}
}

@article{zhang2022structuro,
  title={Structuro-elasto-plasticity model for large deformation of disordered solids},
  author={Zhang, Ge and Xiao, Hongyi and Yang, Entao and Ivancic, Robert JS and Ridout, Sean A and Riggleman, Robert A and Durian, Douglas J and Liu, Andrea J},
  journal={Physical Review Research},
  volume={4},
  number={4},
  pages={043026},
  year={2022},
  publisher={APS}
}

@article{desmarchelier2024topological,
  title={Topological characterization of rearrangements in amorphous solids},
  author={Desmarchelier, Paul and Fajardo, Spencer and Falk, Michael L},
  journal={Physical Review E},
  volume={109},
  number={5},
  pages={L053002},
  year={2024},
  publisher={APS}
}

@article{liu2016driving,
  title={Driving rate dependence of avalanche statistics and shapes at the yielding transition},
  author={Liu, Chen and Ferrero, Ezequiel E and Puosi, Francesco and Barrat, Jean-Louis and Martens, Kirsten},
  journal={Physical review letters},
  volume={116},
  number={6},
  pages={065501},
  year={2016},
  publisher={APS}
}

@article{xu2021atomic,
  title={Atomic nonaffinity as a predictor of plasticity in amorphous solids},
  author={Xu, Bin and Falk, Michael L and Patinet, Sylvain and Guan, Pengfei},
  journal={Physical Review Materials},
  volume={5},
  number={2},
  pages={025603},
  year={2021},
  publisher={APS}
}

@article{szamel2021long,
  title={Long-ranged velocity correlations in dense systems of self-propelled particles},
  author={Szamel, Grzegorz and Flenner, Elijah},
  journal={Europhysics Letters},
  volume={133},
  number={6},
  pages={60002},
  year={2021},
  publisher={IOP Publishing}
}

@article{mandal2020multiple,
  title={Multiple types of aging in active glasses},
  author={Mandal, Rituparno and Sollich, Peter},
  journal={Physical Review Letters},
  volume={125},
  number={21},
  pages={218001},
  year={2020},
  publisher={APS}
}

@article{mandal2021study,
  title={How to study a persistent active glassy system},
  author={Mandal, Rituparno and Sollich, Peter},
  journal={Journal of Physics: Condensed Matter},
  volume={33},
  number={18},
  pages={184001},
  year={2021},
  publisher={IOP Publishing}
}

@article{fruchart2020phase,
  title={Phase transitions in non-reciprocal active systems},
  author={Fruchart, Michel and Hanai, Ryo and Littlewood, Peter B and Vitelli, Vincenzo},
  journal={arXiv preprint arXiv:2003.13176},
  volume={19},
  year={2020},
  publisher={vol}
}

@article{Rossi2023,
	title = {Far-from-equilibrium criticality in the random-field {Ising} model with {Eshelby} interactions},
	volume = {108},
	issn = {2469-9950, 2469-9969},
	url = {https://link.aps.org/doi/10.1103/PhysRevB.108.L220202},
	doi = {10.1103/PhysRevB.108.L220202},
	language = {en},
	number = {22},
	urldate = {2024-08-19},
	journal = {Physical Review B},
	author = {Rossi, Saverio and Biroli, Giulio and Ozawa, Misaki and Tarjus, Gilles},
	month = dec,
	year = {2023},
	pages = {L220202}
}

@article{Ozawa_2023,
  title = {Elasticity, Facilitation, and Dynamic Heterogeneity in Glass-Forming Liquids},
  author = {Ozawa, Misaki and Biroli, Giulio},
  journal = {Phys. Rev. Lett.},
  volume = {130},
  issue = {13},
  pages = {138201},
  numpages = {8},
  year = {2023},
  month = {Mar},
  publisher = {American Physical Society},
  doi = {10.1103/PhysRevLett.130.138201},
  url = {https://link.aps.org/doi/10.1103/PhysRevLett.130.138201}
}

\end{document}